\pdfoutput=1
\documentclass{aa}  

\usepackage{graphicx}
\usepackage{txfonts}
\usepackage{color}
\usepackage{subfigure}
\usepackage{hyperref}
\defcitealias{Mamon2013a}{MBB13}
\defcitealias{DeLucia2007}{DLB07}

\begin{document}

    \title{Dynamical analysis of clusters of galaxies from
    cosmological simulations}

   \author{T. Aguirre Tagliaferro
          \inst{1}
          ,
          A. Biviano\inst{2,3}
          ,
          G. De Lucia\inst{2}
          ,
          E. Munari\inst{2}
          ,
          D. Garcia Lambas \inst{1,4}}

   \institute{Instituto de Astronom\'ia Te\'orica y Experimental, Observatorio Astron\'omico de C\'ordoba,
              Laprida 854, 5000 C\'ordoba, Argentina
    \and
             INAF-Osservatorio Astronomico di Trieste, via G. B. Tiepolo 11, 34131 Trieste, Italy
    \and 
             IFPU-Institute for Fundamental Physics of the Universe, via Beirut 2, 34014 Trieste, Italy
    \and  
            Observatorio Astron\'omico de C\'ordoba, Universidad Nacional de C\'ordoba, C\'ordoba, Argentina 
            }


 
  \abstract
   {Studies of cluster mass and velocity anisotropy profiles are useful tests of dark matter models, and of the assembly history of clusters of galaxies. These studies might be affected by unknown systematics caused by projection effects.
   }
   {We aim at testing observational methods for the determination of mass and velocity anisotropy profiles of clusters of galaxies. Particularly, we focus on the MAMPOSSt technique \citep{Mamon2013a}.}
   {We use results from two semi-analytic models of galaxy formation, coupled with high-resolution N-body cosmological simulations, the catalog of \citet{DeLucia2007} and the FIRE catalog based on the new GAlaxy Evolution and Assembly model.
   We test the reliability of the Jeans equation in recovering the true mass profile when full projected phase-space information is available. We examine the reliability of the MAMPOSSt method in estimating the true mass and velocity anisotropy profiles of the simulated halos, when only projected phase-space information is available, as in observations.}
   {The spherical Jeans equation provides a reliable tool for the determination of cluster mass profiles, also when considering subsamples of tracers separated by galaxy color, except for the central region where deviations may be attributed to dynamical friction effects or galaxy mergers. Results are equally good for prolate and oblate clusters.
   Using only projected phase-space information, MAMPOSSt provides estimates of the mass profile with a standard deviation of $35-69$~\%, and a negative bias of $7-17$\%, nearly independent of radius, and that we attribute to the presence of interlopers in the projected samples. The bias changes sign, that is, the mass is over-estimated, for prolate clusters with their major axis aligned along the line-of-sight.  MAMPOSSt measures the velocity anisotropy profiles accurately in the inner cluster regions, with a slight overestimate in the outer regions, both for the whole sample of observationally-identified cluster members, and, separately, for red and blue galaxies.}
   {}
   
\titlerunning{Dynamics of simulated clusters}
\authorrunning{Tagliaferro et al.}

\maketitle{}
%

\section{Introduction}
Clusters of galaxies, the most massive virialized objects in the Universe, are excellent natural laboratories to study the structure, formation, and evolution of cosmological halos and subhalos. It has been known since a long time that galaxy clusters are dominated by dark matter \citep[DM hereafter,][]{Zwicky33}. Combined gravitational lensing and X-ray  observations of the so-called 'Bullet' cluster provide the strongest evidence that cold, nearly collisionless, dark matter dominates cluster dynamics \citep{Clowe+06b}.

Accurate and precise determination of cluster mass profiles can provide important clues on the properties of dark matter and on astrophysical processes that affect the mass distribution, such as dynamical friction, AGN feedback from the central dominant galaxy, adiabatic contraction, etc. \citep[see, e.g.,][for a review]{Biviano20}. In addition, the determination of cluster mass profiles at large distances from the cluster center can provide a direct estimate of the mass accretion rate \citep{DK14}.

One way to determine a cluster mass profile is from the projected phase-space distribution of member galaxies \citep[e.g.][]{Carlberg+97-equil}, under the assumption of dynamical equilibrium. Other methods are based on the gravitational lensing distortions from background galaxies and the assumption of hydrostatic equilibrium of the intracluster gas \citep[e.g.][]{Allen98,EDGM02,HYG04,Pratt+19}. Given the cluster mass distribution, it is possible to determine the orbital distribution of cluster members \citep[e.g.][]{NK96}. As galaxies enter into clusters through hierarchical accretion, their orbits keep important information on the processes that lead to the cluster mass assembly and internal dynamical relaxation \citep[e.g.][]{Biviano2004}. Moreover, knowledge of the orbits of cluster galaxies is crucial to understand environmental effects that differentiate galaxy evolution in clusters relative to the field \citep[e.g.][]{Lotz+19,Tonnesen19,Joshi+20}.

To test the reliability of the methods of cluster mass profile determination, one can rely on the analysis of halos extracted from cosmological numerical simulations of a $\Lambda$CDM model. Simple analytical models cannot capture the full complexity of highly non-linear dynamics and the astrophysical effects of the intra-cluster medium. Semi-analytical models (\citealt{Cole1991, White&Frenk1991}) allow us to combine our understanding of the astrophysical processes at work in clusters of galaxies with numerical simulations, thus presenting a successful approach to the study of the dynamical and astrophysical processes that shape cluster mass distributions and the orbits of cluster galaxies.

An example of the use of cosmological simulations to test cluster mass determination came from \cite{Biviano2006}. By simulating the observational procedure, in particular taking into account projection effects, \citet{Biviano2006} found that the line-of-sight velocity dispersion ($\sigma_{\rm{los}}$) provides more accurate cluster mass estimates than the virial theorem. 
\cite{Munari2013} analysed the relation between the masses of cluster- and group-sized halos, extracted from $\Lambda$CDM cosmological N-body and hydrodynamic simulations, and their halos velocity dispersions at different redshifts. They found that using DM particles as tracers of the gravitational potential leads to a relation between the 3D velocity dispersion $\sigma_v$ and the halo mass $M$, $\sigma_v \propto M^{\alpha}$ with $\alpha \simeq 1/3$, as expected theoretically, from the virial scaling \citep{Evrard+08}.
 \cite{Munari2013} also found that the $M-\sigma_v$ relation is steeper when using subhalos and galaxies as tracers,  $\alpha > 1/3$, possibly because of dynamical friction and tidal disruption processes. 
\citet{SMBD13} investigated the Millennium numerical simulation coupled with the semi-analytical model of \citet[][DLB07 hereafter]{DeLucia2007}, to predict the scatter in the $M-\sigma_{\rm{los}}$ relation, and identified the main source of this scatter in the triaxiality of the velocity ellipsoid. The presence of interlopers (misidentified non-cluster members) is the main source of scatter in the $M-\sigma_{\rm{los}}$ relation, and the scatter increases dramatically for samples of less than 30 galaxies. Restricting the selection of cluster members to red galaxies reduces the fraction of interlopers and therefore the scatter in $M-\sigma_{\rm{los}}$ relation. However, dynamical friction decelerates the brightest (red) galaxies, producing a biased estimate of $\sigma_{\rm{los}}$.

\citet{Old+15} used mock clusters from numerical simulations to compare the relative performance of 25 different methods for cluster mass determinations, based on the spatial and/or velocity distributions of galaxies or on the cluster richness. They found a wide range of rms errors in the estimate of $\log M_{200}$\footnote{$M_{200}$ is the mass contained within a sphere of radius $r_{200}$, that encloses a mean overdensity 200 times the critical density of the Universe at the redshift of the halo.} from the 25 different methods, ranging from 0.2 to 1.1 dex. Their study was then extended by \citet{Old+18} to assess the importance of dynamical substructure on cluster mass estimates based on the projected phase-space distribution of galaxies. They found that cluster masses tend to be systematically over-estimated for clusters of low masses.

Using simulated galaxies in halos extracted from cosmological simulations,  in this paper we extend the scope of previous investigations, and test not only the accuracy of cluster mass estimates, but also the accuracy of mass {\em profile} estimates. Using cluster-size halos extracted from numerical simulations coupled with semi-analytical models, we consider both estimates based on full 6D phase-space distribution of tracers of the gravitational potential, as well as estimates obtained from projected phase-space distributions. In particular, we test the capability of the Modeling Anisotropy and Mass Profiles of Observed Spherical Systems algorithm \citep[MAMPOSSt,][MBB13 hereafter]{Mamon2013a} to recover the mass profile from projected phase-space information. In addition, we also investigate the velocity-anisotropy profiles of cluster-size halos, and how well can MAMPOSSt reconstruct these profiles from the limited projected phase-space information.

This paper is structured as follows. In Sect.~\ref{s:samp}, we describe the semi-analytic models used for this work and define the samples used in our analysis.  In Sect.~\ref{s:mr}, we determine the mass profiles using the full 6D phase-space information (Sect.~\ref{ss:mr6d}) and the 3D projected phase-space information only (Sect.~\ref{ss:mr3d}). In Sect.~\ref{s:beta}, we determine the velocity anisotropy profiles, using the full 6D phase-space information (Sect.~\ref{ss:beta6d}) and the 3D projected phase-space information only (Sect.~\ref{ss:betaproj}). We discuss our results in Sect.~\ref{s:disc} and draw our conclusions in Sect.~\ref{s:conc}.

\section{The samples of galaxy clusters}
\label{s:samp}
\subsection{Dark matter particle simulation}
We use the dark matter cosmological N-body Millennium Simulation \citep[MI hereafter,][]{Springel2005}. This traces the evolution of $2160^3$ DM particles of mass $8.6 \times 10^{8} h^{-1} M_{\odot}$ within a comoving box of size 500 $h^{-1}$ Mpc on a side, from redshift $z=127$ to $z=0$. 
The cosmological parameters of the simulation are: 
$\Omega_{m}$ = 0.25, $\Omega_{\Lambda}$ = 0.75, $\Omega_b = 0.045$, $h = 0.73$, $n = 1$, and $\sigma_8 = 0.9$, with the Hubble constant parametrized as $H_0 = 100 \, h \, \mathrm{km} \, \mathrm{s}^{-1} \mathrm{Mpc}^{-1}$. The resolution of the simulation allows galaxies down to a stellar mass of $\sim 10^{9}\,h^{-1}\,{\rm M}_{\odot}$ to be resolved. The large volume of the simulation allows us to identify a statistical sample of galaxy clusters for the analysis presented below.  

\subsection{Semi-analytic models of galaxy formation}
\label{ss:sam}
In the currently accepted paradigm for structure formation, galaxies form through the condensation of gas at the center of DM halos, that evolve hierarchically: small halos form first and later merge to form more and more massive systems. Galaxy evolution is driven by several astrophysical processes including gas cooling, star formation, stellar and AGN feedback, and dynamical processes such as dynamical friction and tidal interactions. Semi-analytic models (SAMs) represent a powerful tool to include in numerical simulations the relevant astrophysical processes. The approach consists in modelling these processes using analytical or numerical prescriptions based on observational and/or theoretical results. In this way, the process of galaxy formation and evolution is expressed using a set of coupled differential equations that allow the user to compute the flow of baryons between the different components of model galaxies (e.g. hot gas, cold gas, ejected reservoir, and stars). Since the computational costs of the method are rather limited, it allows an efficient investigation of the parameter space and of the influence of different specific assumptions. 

In this paper, we analyze the $z = 0$ outputs from two different SAMs both based on the MI. Specifically, we use outputs from the original model described in \citetalias{DeLucia2007} and from the new GAlaxy Evolution and Assembly (GAEA, hereafter) model published in \citet{Hirsch2016}. The latter is an evolution of the former model including several major updates. In particular, the version of GAEA used in this work includes (i) a scheme to account for the non-instantaneous recycling of gas, metals and energy \citep{DeLucia2014}, that allows individual metal abundances to be traced in detail, and (ii) a stellar feedback prescription that is partly based on results from numerical simulations and that, in the framework of GAEA, allows the evolution of the galaxy stellar mass function to be reproduced up to $z\sim 3$ \citep{Hirsch2016}. With respect to the DLB07 model, GAEA provides a better overall agreement with observational data. In particular, it solves the excess of low-to-intermediate mass galaxies that has been discussed in several previous studies \citep[see, e.g.,][and references therein]{Fontanot2009}; it provides a better agreement with the observed fractions of passive galaxies as a function of galaxy stellar mass \citep{Hirsch2016,DeLucia2019}; and it reproduces relatively well the observed evolution of the relation between galaxy stellar mass and gaseous metallicity \citep{Hirsch2016,Xie2017}. We refer to the original papers for full details on the modelling adopted. The updated model is not without problems: in previous work, we have shown that the most massive galaxies are characterized by levels of activity that are larger than typically observed, and that the colour bimodality is not as well pronounced as in observational data for intermediate mass galaxies. We consider that working with two different SAMs that are both based on the MI can provide information about how results may (or may not) depend on the physical prescriptions adopted.

For the analysis below, we have selected 100 halos with $M_{200}>10^{14} M_{\odot}$ at $z=0$ from the simulation box. The halos were selected randomly with a uniform distribution in different mass bins. 
The DLB07 and GAEA galaxy clusters catalogs are contained in cubic boxes $5 \, r_{200}$ on a side. 

\subsection{Projected phase-space samples}
\label{ss:pps}
Observers have only access to projected phase-space distributions of galaxies in clusters, namely projected distances from the cluster center and rest-frame line-of-sight velocities. To mimic the observational data, we project the DLB07 and GAEA clusters along three orthogonal planes. We restrict the projected samples to galaxies within 3 Mpc from the cluster center in projection. We adopt the cluster centers as given in the simulation. We do not try to mimic the observational uncertainty in the cluster center definition. This uncertainty is typically $\sim 50$ kpc \citep{AMKB98}, so it does not affect our dynamical analysis that excludes the central 50 kpc. Also, we adopt the cluster mean velocities as given in the simulation to define the rest-frame velocities of its galaxies. The typical observational error on the mean cluster velocity is $< 50$ km~s$^{-1}$ when $>100$ cluster members are available \citep[][]{Biviano2013,MBM14}, so we can neglect it since it is small enough not to affect the present dynamical analysis.

\begin{figure}[ht]
    \centering
    \includegraphics[height=7cm]{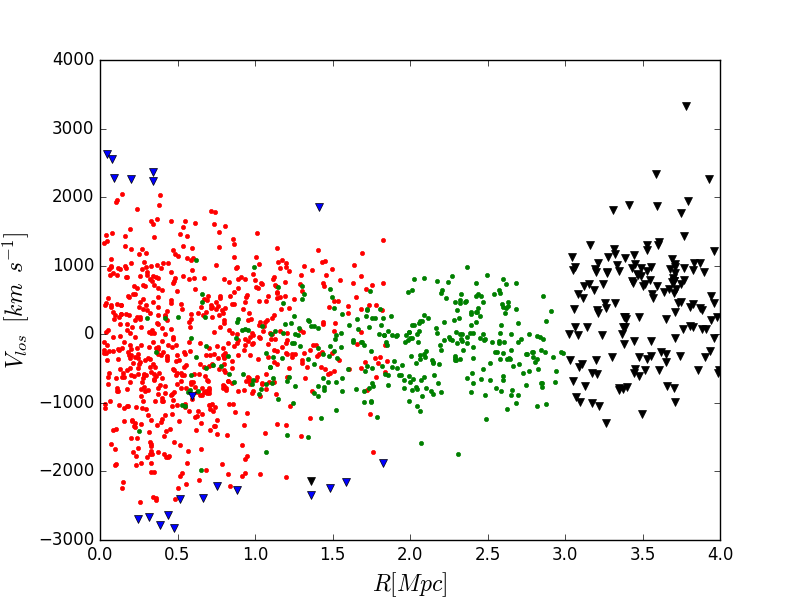}
    \caption{Projected phase-space diagram (line-of-sight velocities vs. cluster-centric distances) of a simulated cluster from the DLB07 sample. Dots (triangles) are galaxies selected (respectively, not selected) by CLEAN as cluster members.  Red dots are real members, located within the $r_{200}$ sphere, while green dots are interlopers, galaxies identified as members but outside the $r_{200}$ sphere. Blue triangles are real members (within the $r_{200}$ sphere) incorrectly rejected as interlopers by CLEAN, while black triangles are galaxies beyond the 3 Mpc limiting distance of our projected-phase space dynamical analysis.}
    \label{fig:phase_space}
\end{figure}

The projected data-sets do not contain only cluster members, but also galaxies located in the 3 Mpc circular projected region around the cluster center, but outside the 3D virial sphere. We call these galaxies 'interlopers'. To mimic the observational approach, we use the Clean method \citep{Mamon2013a} to try removing these interlopers from the sample of cluster members.
The Clean method uses a robust estimate of the cluster line-of-sight velocity dispersion, $\sigma_{{\rm los}}$, to guess the cluster mass using a scaling relation. It then adopts a NFW profile \citep{Navarro1997}, the theoretical concentration-mass relation of \cite{Maccio2008} and the velocity anisotropy profile model of \cite{Mamon2010}, to predict $\sigma_{{\rm los}}(R)$ and to remove from the sample of cluster members all galaxies with projected rest frame absolute velocities $>2.7\sigma_{{\rm los}}$ at any radius $R$, in an iterative fashion. In Fig. \ref{fig:phase_space} we show an example of Clean membership selection in the projected phase-space diagram  of a cluster of the DLB07 sample. 

We call 'C' the sample composed by members identified by the Clean algorithm, and 'RM' (for 'real members') the sample composed by galaxies contained in the 3D virial sphere. In addition, to simulate the observational case of poor sampling statistics, we also consider an analog of the 'C' sample in which the Clean algorithm is run on a subset of 100 galaxies extracted at random from each 3 Mpc projected cluster region. We call this sample 'C100'.

\section{Mass profiles}
\label{s:mr}
\subsection{Full phase-space}\label{6Dmass}
\label{ss:mr6d}

The collisionless Boltzmann equation describes the density of tracers of the gravitational potential as a function of position, velocity and time, that is, the phase-space density $f(r,v,t)$. One generally assumes that a cluster of galaxies is stationary, so that the phase-space density $f$ does not depend on time. Further assumptions are spherically symmetry, and equality of the two components of the velocity dispersion in the tangential directions of the galaxies motion, $\sigma_{\theta}\equiv\sigma_{\phi}$. 
With these assumption the spherical Jeans equation follows from the Boltzmann equation,
\begin{equation}
    M_J(r) = -\frac{r\sigma_{r}^2}{G} \bigg[ \frac{d \textrm{ln} \nu}{d \textrm{ln} r}+\frac{d\textrm{ln}\sigma_r^2}{d\textrm{ln}r}+2\beta(r) \bigg]
    \label{eq:jeans}
\end{equation}
where $M_J(r)$ is the mass within a radius $r$, $G$ is the gravitational constant, $\nu(r)$ is the number density profile of the tracers of the gravitational potential, and $\beta(r)$ is the velocity anisotropy profile, 
\begin{equation}
    \beta \equiv 1 - (\sigma_{\theta}/\sigma_{r})^2,
    \label{eq:beta}
\end{equation}
where 
$\sigma_r$ is the radial component of the velocity dispersion. Purely radial orbits of the tracers of the gravitational potential correspond to $\beta(r)=1$, while purely tangential orbits correspond to $\beta \rightarrow -\infty$.

\begin{figure}[ht]
    \includegraphics[height=7cm]{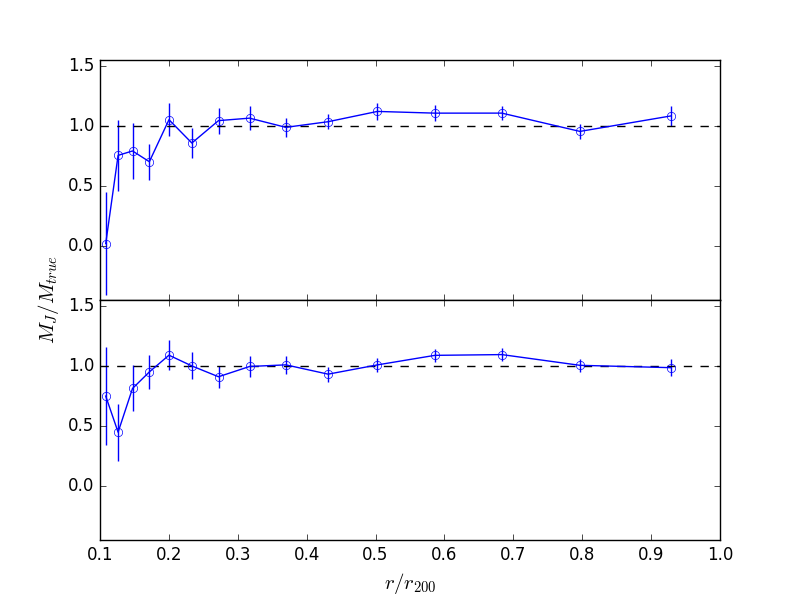}
    \caption{Median of $M_J(r)/M_{true}$ for 100 clusters (blue solid line) and its uncertainty $\sigma/\sqrt{N}$. Top panel: clusters from the DLB07 sample. Bottom panel: clusters from the GAEA sample.}
    \label{fig:MJMtrue_ratio}
\end{figure}

Using full phase-space information for true cluster members 
we estimate $M_J(r)$ through the Jeans equation (Eq. \ref{eq:jeans}) for each cluster of the DLB07 and GAEA data-sets. We also estimate the true mass profile $M_{true}$,
evaluated by direct sum of the masses of the particles within each given radius $r$. 
Fig. \ref{fig:MJMtrue_ratio} shows the ratio of the median $M_J(r)$ for the 100 DLB07 clusters (upper panel) and for the 100 GAEA clusters (lower panel) and the median $M_{true}$. The confidence interval on the mass profile ratio is given by $\sigma/\sqrt{N}$ where $\sigma$ is the rms of all the $N=100$ $M_J(r)$. To compute the terms showed in Eq. \ref{eq:jeans} we use binning. The number of bins we choose, 15, is a compromise between the requirements of good spatial resolution and sufficient number statistics.
Perfect agreement between $M_{true}$ and $M_J$ is expected if all the above mentioned assumptions that validate the Jeans equation are verified. $M_J \approx M_{true}$ at all radii $r \gtrsim 0.2 r_{200}$. $M_J \lesssim M_{true}$ at smaller radii; this difference, albeit not very significant, can be due to dynamical friction, invalidating the assumption of a collisionless fluid and thereby the applicability of the Jeans equation. 

Overall this analysis indicates that the dynamics of clusters in the DLB07 and GAEA samples satisfies the Jeans equation eq.~(\ref{eq:jeans}). Our finding supports and extends the results of \citet{Biviano2016} who found that the internal dynamics of cluster-size halos extracted from a cosmological simulation satisfies the virial theorem, that is derived from the Jeans equation via integration \citep{Binney1987}.

\subsubsection{Red and blue galaxies}
\label{sss:6dcolors}
The good agreement between $M_{true}(r)$ and $M_J(r)$ (see Sect.~\ref{ss:mr6d}) indicates that clusters in the DLB07 and GAEA samples are close to dynamical equilibrium at $z=0$. 
While this might be the case when all galaxies are considered, different samples of galaxies that entered the cluster at different times might show deviations from dynamical equilibrium.

\begin{figure}[ht]
    \includegraphics[height=7cm]{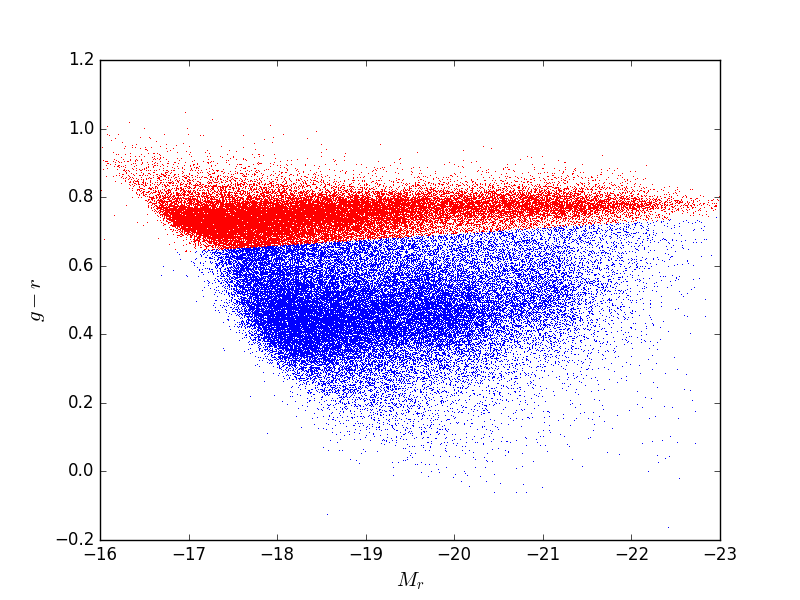}
    \caption{Color-magnitude diagram for one cluster from the DLB07 sample. Red and blue dots correspond to selected red and blue galaxies.}
    \label{fig:color-mag}
\end{figure}

\begin{figure}[ht]
    \includegraphics[height=7cm]{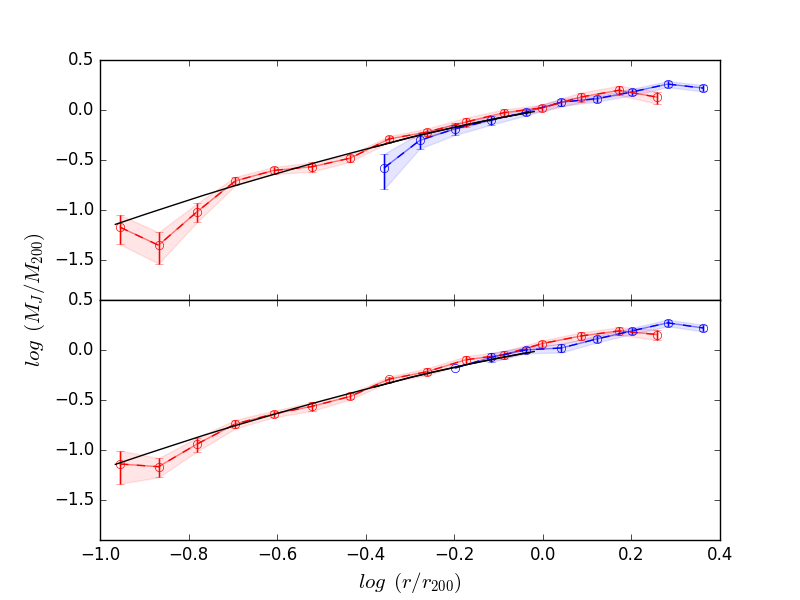}
    \caption{Median of 100 cluster $M_{J}$ and its uncertainty obtained using red galaxies (red points with error bars and line) or blue galaxies (blue points with error bars and line) as tracers of the gravitational potential, compared to to $M_{true}(r)$ (black line). Top panel: clusters from the DLB07 sample. Bottom panel: clusters from the GAEA sample.}
    \label{fig:6dcolor}
\end{figure}

Using the $g-r$ color-magnitude diagram, we split the clusters samples into two different populations: red and blue galaxies. We show an example for one DLB07 cluster in Fig. \ref{fig:color-mag}, where red and blue dots correspond to the selected red and blue galaxies respectively. In Fig. \ref{fig:6dcolor} we show the median of 100 cluster $M_J(r)$ obtained using only blue (blue lines) and only red (red lines) galaxies as tracers of the gravitational potential. These median mass profiles are compared to the median $M_{true}(r)$ (black line). The top and bottom panel correspond to the DLB07 and GAEA samples, respectively. The different radial range spanned by $M_J(r)$ when using the red and blue galaxies as tracers, is related to the spatial segregation of cluster galaxies with color, red galaxies residing closer to the cluster center than blue galaxies. This different spatial distribution of red and blue galaxies does not seem to hamper the determination of $M_J(r)$ that remains pretty close to $M_{true}(r)$ independently of the population chosen. This finding is consistent with the observational results of \citet{Carlberg+97-equil} who were the first to recognize that applying the Jeans equation to the red and blue galaxy subsamples separately gives statistically consistent cluster mass profiles. 

\subsubsection{Prolate and oblate clusters}
\label{sss:shapes}
The good agreement between $M_{true}(r)$ and $M_J(r)$ suggests that the spherical assumption adopted for the Jeans equation eq.~(\ref{eq:jeans}) is acceptable. However, clusters and groups of galaxies are known to be aspherical \citep{Limousin2013} and mostly prolate \citep{Wojtak2013}. Numerical simulations indicate that more massive dark matter halos tend to be more prolate and aspherical than less massive halos \citep{Kasun2005,Paz2006}, probably as a result of the accretion process. 

\begin{figure}[ht]
    \includegraphics[height=5.5cm]{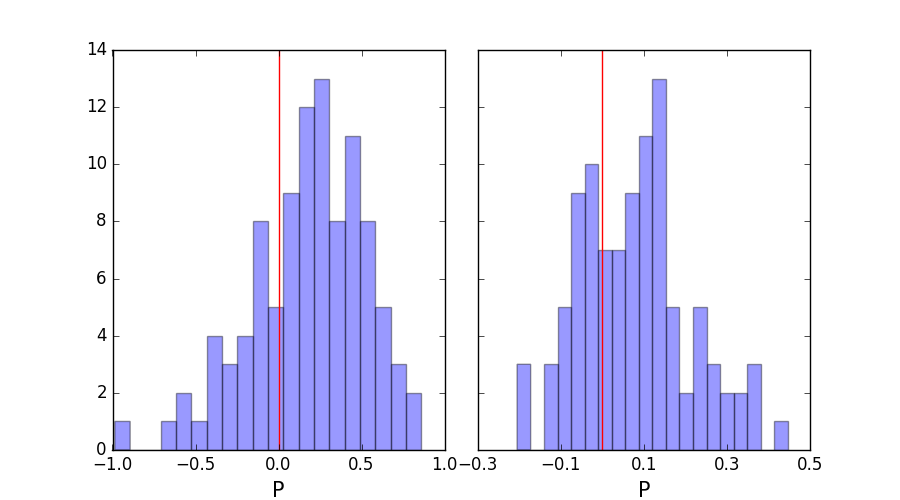}
    \caption{$P$ value distribution of the clusters in our samples. The vertical red line indicate $P=0$. Oblate (resp. prolate) shape have $P<0$ (resp. $P>0$). 
    Left panel: DLB07 sample. Right panel: GAEA sample.}
    \label{fig:triaxial}
\end{figure}

For each cluster we evaluate the inertia tensor \citep{Paz2006}, 
\begin{equation}
    I_{ij}=(1/N_h) \, \Sigma_{\alpha=1}^{N_h}X_{\alpha i}X_{\alpha j},
    \label{eq:inertia}
\end{equation}
where $X_{\alpha i}$ is the \textit{i}-th component of the displacement vector of a particle $\alpha$ relative to center of mass, and $N_h$ is the number of particles in the halo. The matrix eigenvalues correspond to the square of the semi-axis ($a, b, c$ where $a > b > c$) of the characteristic ellipsoid that best describes the spatial distribution of the halo members.
We calculate for each cluster the inertia tensor using the positions of the halo members with $r<1.5 \, r_{200}$. We then evaluate the triaxiality parameter $P= \ln(ca/b^2)$.
For an oblate ellipsoid $P < 0$ while for a prolate ellipsoid $P > 0$. In Fig.~\ref{fig:triaxial} we show the distribution of $P$ values for clusters of the DLB07 and GAEA samples (left and right panels, respectively).
Slightly more than 2/3 of the clusters (74\% for the DLB07 sample, 67\% for the GAEA sample) have a prolate shape. The excess of prolate clusters we find is much larger than found by \citet{SMBD13} on the same DLB07 data-set, but their definition of prolateness  is based on the ratio of different components of the cluster velocity dispersion, while our definition is based to the spatial distribution of galaxies.

\begin{figure}[ht]
    \includegraphics[height=7cm]{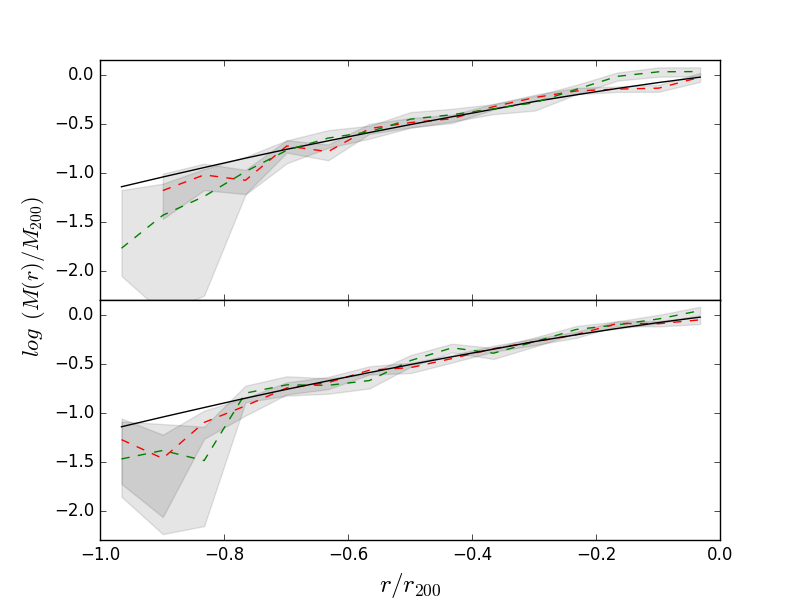}
    \caption{Mass profiles for prolate and oblate clusters. The median $M_{true}(r)$ of all clusters is represented by the black solid curve, the median $M_{J}(r)$ for prolate clusters is represented by the dashed red curve and the median $M_{J}(r)$ for oblate clusters by dashed green curve. Shaded regions represent error bars $\sigma/\sqrt{N}$. Top panel: DLB07 sample. Bottom panel: GAEA sample.}
    \label{fig:Mrshape}
\end{figure}

To test the reliability of the Jeans equation in systems that deviate from spherical symmetry we compute $M_{J}(r)$ separately for prolate and oblate clusters. Results for both samples are shown in Fig. \ref{fig:Mrshape} (top panel: clusters in the DLB07 sample, bottom panel: clusters in the GAEA sample). Black solid lines represent the median $M_{true}(r)$ for all clusters. Red dashed lines represent the median $M_{J}(r)$ for prolate clusters while green dashed lines represent the median $M_{J}(r)$ for oblate clusters. Outside the central regions, the agreement between $M_{true}(r)$ and $M_{J}(r)$ is equally good for prolate and oblate clusters. 

We conclude that cluster asphericity does not hamper the applicability of the Jeans equation if all relevant terms in the Jeans equation are calculated in spherical shells. Our conclusion is valid on an average sense, and as such it confirms the result obtained by \citet{vanderMarel+00} for a stack of real clusters. Our result is also consistent with the conclusion by \citet{SMBD13} that the mild excess of prolate with respect to oblate clusters has little effect on dynamical mass estimates, in a statistical sense. 

\subsection{Projected phase-space}
\label{ss:mr3d}
So far we have considered the full phase-space information for cluster mass profile determination (Sect.~\ref{6Dmass}). However, observers only have access to projected phase-space information in the case of clusters of galaxies, that is, projected distances from the cluster center, and line-of-sight velocities. Here we consider how well can we reproduce $M_{true}(r)$ from the limited information available from projected phase-space.

\subsubsection{The MAMPOSSt method}
\label{sss:mamp}
To solve the Jeans eq.~(\ref{eq:jeans}) for $M_J(r)$, knowledge of the 3D profiles  $\nu(r), \sigma_r(r)$ and $\beta(r)$ is required. The 3D number density profile $\nu(r)$, can be derived from the projected number density profile, $N(R)$, an observable, using the Abel inversion equation \citep{Binney1987}. Unfortunately, $\sigma_r$ cannot be derived from the other observable, $\sigma_{\rm{los}}$, without knowledge of $\beta(r)$, and as a result the solution for $M_J(r)$ is degenerate with the solution for $\beta(r)$. 

MAMPOSSt \citepalias{Mamon2013a} attempts to solve this degeneracy by using the full available information in projected phase-space, following an early suggestion by \citet{Merritt87}. MAMPOSSt adopts models for $M_J(r), \nu(r)$, and $\beta(r)$ with any chosen number of free parameters, to assess the probability of observing a galaxy member of a cluster, at a given position in projected phase-space. The best-fit parameters of the adopted models are determined by maximizing the product of the probabilities of all cluster members. Since $\nu(r)$ can be directly determined from the projected number density profile of the tracer, $N(R)$, using Abel's inversion equation, MAMPOSSt is generally used in its so-called Split mode, where $\nu(r)$ is determined independently from a maximum likelihood fit of suitable models to $N(R)$, and MAMPOSSt is left the task of determining the best-fit parameters of the $M_J(r)$ and $\beta(r)$ models only. 

MAMPOSSt has been applied to several galaxy cluster samples \citep[e.g.][]{Biviano2013,Biviano2016,Biviano+17a,Capasso+19,Mamon+19}. It has also been tested on cluster-size halos from cosmological simulations. Using a set of 11 halos \citetalias{Mamon2013a} estimated the bias and inefficiency of MAMPOSSt in the recovery of the virial radius, tracer scale radius, dark matter scale radius and velocity anisotropy. \citet{Old+15} run a challenge to determine the mass of 1000 mock clusters and MAMPOSSt was the second most efficient among the 23 algorithms that took part to the challenge. In our analysis we extend the testing of MAMPOSSt to measure its efficiency in recovering the whole mass and velocity anisotropy profiles of simulated clusters.

We choose to model $M_J(r)$ using the NFW profile \citep{Navarro1997}:
\begin{equation}
    M_{NFW}(r) = M_{200}\frac{\ln(1+r/r_{-2})-r/r_{-2}(1+r/r_{-2})^{-1}}{\ln(1+c)-c/(1+c)},
    \label{eq:nfw}
\end{equation}
 where $r_{-2}$ is the scale radius where the logarithmic derivative of the mass density profile equals $-2$ and $c \equiv r_{200}/r_{-2}$ is the mass profile concentration.

We choose to adopt a simplified version of the \citet{Tiret+07} profile to model $\beta(r)$ (Tiret model hereafter),
\begin{equation}
    \beta(r)=\beta_{\infty}\frac{r}{r+r_{-2}}
    \label{eq:betatiret}
\end{equation}
where $\beta_{\infty}$ is the value of velocity anisotropy value at very large radii. In the Tiret model the velocity distribution is isotropic at the center, and becomes increasingly more radial outside. This model provides a good fit to the velocity anisotropy profiles for cosmological cluster-mass halos \citep{Mamon2013a}. We discuss the effects of generalizing this profile in Sect.~\ref{ss:betaproj}.

\subsubsection{Results}
\label{sss:mr3dres}
\begin{figure}[ht]
    \centering
    \subfigure{\includegraphics[width=0.23\textwidth,height=4.5cm]{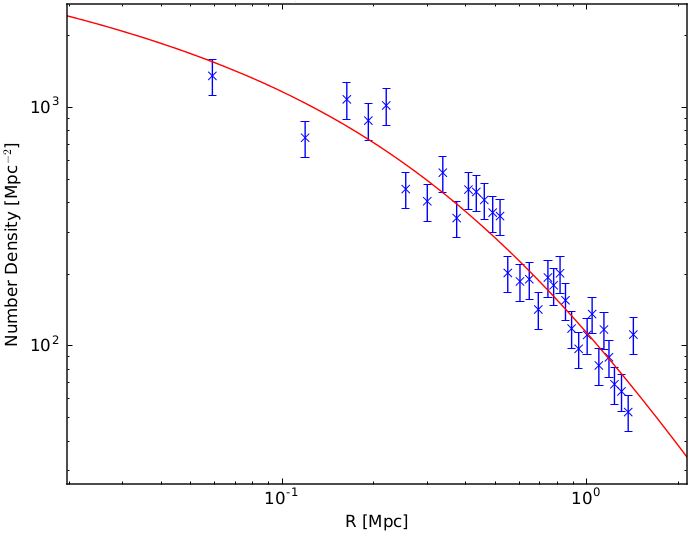}\hspace{0.2mm}}
    \subfigure{\includegraphics[width=0.23\textwidth,height=4.5cm]{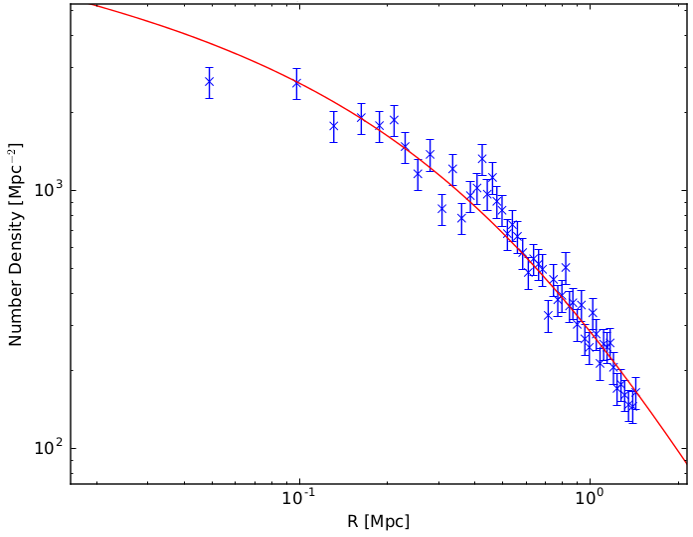}}
    \caption{Example of the projected NFW profile fit (solid red line) and to the $N(R)$ (points with 1 $\sigma$ error bars) for a cluster from the DLB07 (left panel) and GAEA (right panel) sample.}
    \label{fig:pNFW}
\end{figure}

We run MAMPOSSt in Split mode, as described in Sect. \ref{sss:mamp}, and 
we fit each cluster $N(R)$ with a projected NFW profile \citep{Bartelmann1996}. In Fig.~\ref{fig:pNFW} we show two examples of such fits. The fit is performed with a maximum likelihood technique, and the only free parameter of the model is the scale radius of the NFW profile, $r_{\nu}$, since the normalization of the NFW model is fixed by the requirement that the integral of the probability distribution of the model is equal to the number of observed galaxies.

Using the best-fit parameter of the fit to $N(R)$ we then run MAMPOSSt to find the maximum-likelihood fit to the radially-dependent distribution of galaxy velocities in each cluster of the DLB07 and GAEA samples. The free parameters in the MAMPOSSt analysis are, for $M_{NFW}(r)$, ${r_{200}, r_{-2}}$, or, equivalently, ${M_{200}, c_{200}}$, and for $\beta(r)$, $\beta_{\infty}$. We consider the three data-sets described in Sect.~\ref{ss:pps}, namely C, C100, and RM.

In Table~\ref{tab:m200} we list the average of the 300 (3 projections $\times 100$ clusters) ratios of the MAMPOSSt estimates and true values of $M_{200}$, for the three data-sets, C, C100, and RM. MAMPOSSt appears to slightly underestimate the true $M_{200}$ values when members are identified with the Clean procedure (see Sect.~\ref{ss:pps}, and to slightly overestimate the true $M_{200}$ values when only true members (i.e. galaxies within the $r_{200}$ sphere) are considered. We measure a bias in the estimate of 
$M_{200}$ of 7--15 per cent and a standard deviation of 35--69 per cent for the C sample. For comparison, \citetalias{Mamon2013a} measured a bias of 6 (4) per cent and a standard deviation of 42 (respectively, 30) per cent for their run with Clean membership selection, the NFW and Tiret models, and 100 (respectively 500) tracers per cluster. Presumably their results were too optimistic due to the very limited number of halos analysed (11 in total).

Interestingly, the results we find are not much worse for the C100 sample than for the C sample, suggesting that the bias and inefficiency of MAMPOSSt estimates are not so much related to statistical errors but to systematic or intrinsic uncertainties. The bias seems to be due to a systematic error related to the selection of cluster members. In fact, the mass bias changes sign when the RM sample is considered. On the other hand, the standard deviation of the RM sample is very similar to that of the C sample, and this suggests that the origin of the standard deviation is intrinsic, perhaps related to the unrelaxed dynamical state of a fraction of clusters, as suggested by the previous analyses of \citet{Biviano2016} and \citet{Old+18}. 

\begin{figure}[ht]
    \includegraphics[height=7cm]{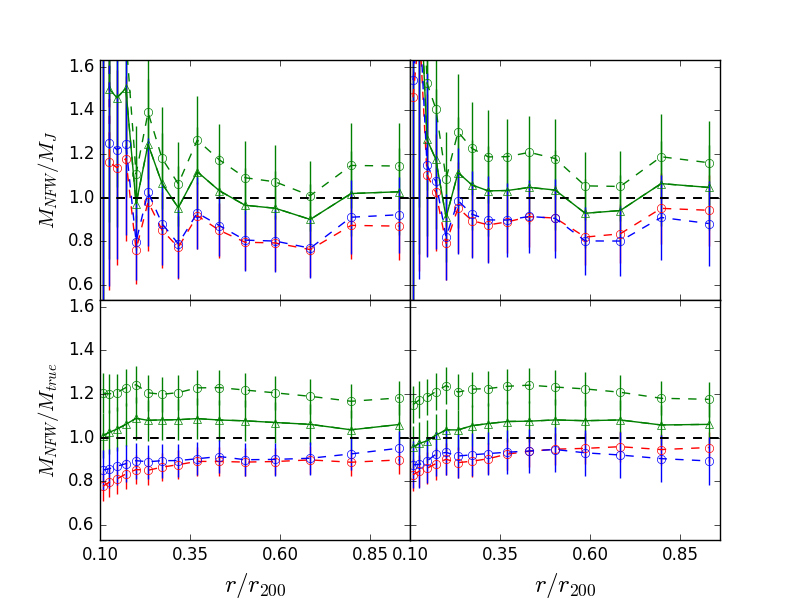}
    \caption{Ratio of different mass profile determinations. Top panels: ratio between the median NFW mass profile determined with MAMPOSSt, $M_{NFW}(r)$ and based on projected phase-space information, and the mass profile determined by direct application of the Jeans equation using full phase-space information, $M_{Jeans}$. Red, blue and green symbols refer to the C, C100, and RM samples, respectively. The solid line represents the result for the RM sample obtained by adopting $r_{200}$ as the maximum integration distance in MAMPOSSt.
    Bottom panels: ratio between $M_{NFW}(r)$ and the true mass profile, $M_{true}$. Left, respectively right, panels display results obtained on the DLB07, respectively GAEA, samples.}
    \label{fig:mass_ratio}
\end{figure}

In Fig. \ref{fig:mass_ratio}
we show two NFW median mass profiles, evaluated using the MAMPOSSt maximum likelihood fit values of $r_{200}$ and $r_{-2}$ for the DLB07 and GAEA samples, and compare them with both $M_{J}(r)$ and $M_{true}(r)$ (see Sect. \ref{ss:mr6d}). We notice that the mass bias we have observed is not specific to $r=r_{200}$, but is present at all radii. The mass bias is similar for $M_{NFW}/M_{J}$ and $M_{NFW}/M_{true}$ but more significant for the latter. Given that MAMPOSSt uses the Jeans equation, it is not surprising that $M_{NFW}(r)$ is closer to $M_{J}(r)$ than to $M_{true}$. 

We showed in Sect.~\ref{ss:mr6d} that the Jeans equation provides an unbiased mass estimate when using 6D phase-space information, so it is surprising that MAMPOSSt over-estimates the true mass when using the true cluster members. We think this is related to the fact we adopted the most commonly accepted definition of "true cluster members", that is galaxies within the virial sphere \citep[other definitions are possible, see, e.g.][]{Wojtak+07}. Restricting the sample of true members to the virial sphere artificially forces MAMPOSSt integrals of the 3D radius to zero beyond $r_{200}$. As a consequence, the mass normalization has to increase to compensate for the lost contributions to the integral outside $r_{200}$ \citep[see eq.~(27) in][]{Mamon2013a}. In other terms, the maximum line-of-sight distance of RM galaxies is $r_{200}$ and it differs from that adopted in MAMPOSSt \citep[$\infty$, see eq.(27) in][]{Mamon2013a}. To confirm the origin of the mass profile bias for the RM sample, we changed
the integration limit of MAMPOSSt from $\infty$ to $r_{200}$. After this change, the mass estimation bias for the RM sample disappears (see Table~\ref{tab:m200}, "RM ($r_{200}$)" line, and Fig.~\ref{fig:mass_ratio}, solid green curve).

The underestimate of $M_{200}$ for the C and C100 samples is probably related to remaining interlopers in the cluster sample after the Clean procedure has been applied. These interlopers are for the most part galaxies far away from the cluster center \citep{MBM10} with relatively small velocities relative to the cluster members that are at the same projected distance, but much closer to the cluster center in real space. The small velocities of these interlopers make it impossible for Clean to reject them as non-members (see Fig.~\ref{fig:phase_space}), and, at the same time, decrease the amplitude of the observed velocity distribution of the cluster \citep[see][]{Cen1997,Biviano2006}, ultimately leading to a decrease in the estimate of $M_{200}$. The mass bias appears to be somewhat larger for the C100 data-set than for the C data-set, suggesting that smaller data-sets are more likely to be contaminated by non-members in the Clean procedure. 

\begin{table}[]
\centering
\caption{MAMPOSSt vs. true $M_{200}$ ratio}
\label{tab:m200}
    \begin{tabular}{lcc}
    \hline
Data-set    &   \multicolumn{2}{c}{$M_{200_{MAM}}/M_{200}$} \\
            & DLB07           &     GAEA         \\ 
    \hline 
C                   &  $0.85 \pm 0.04$  &  $0.93 \pm 0.02$   \\ 
C100                &  $0.84 \pm 0.04$  &  $0.83 \pm 0.03$   \\
RM                  &  $1.10 \pm 0.03$  &  $1.12 \pm 0.03$   \\ 
RM ($r_{200}$)      &  $1.01 \pm 0.02$  &  $1.03 \pm 0.03$   \\
    \hline
    \end{tabular}
    \tablefoot{$M_{200_{MAM}}/M_{200}$ is the average of the ratios (300 values $=3$ projections $\times 100$ clusters per sample) between the MAMPOSSt determined value of $M_{200}$ and the corresponding true value. Errors are $\sigma/\sqrt{N}$.}
\end{table}

\subsubsection{Line-of-sight alignment}
\label{sss:shapes_aligned}
We showed in Sect.~\ref{sss:shapes} that the spherical Jeans equation can be successfully used to estimate a cluster mass profile even if the cluster is not spherically symmetric. However, cluster shapes have an impact on mass estimates because of projection effects. Mass estimates obtained from X-ray data or gravitational lensing are known to be biased high when a cluster is observed with its major axis aligned along the line-of-sight \citep{Rasia+13,Meneghetti+14}. The same is also true when masses are estimated from the kinematics of cluster galaxies \citepalias{Mamon2013a}, because velocity ellipsoids are aligned with the halo major axis \citep{WGK13}. For this reason we here consider the effect of line-of-alignment of a cluster major axis. 

The effect of major axis line-of-sight alignment on a cluster mass estimate is expected to be stronger for prolate clusters. We therefore consider only the 74 DLB07 and 67 GAEA prolate clusters for this analysis. We split each of these samples into two subsamples of clusters that have their major axis preferentially aligned with or orthogonal to the line-of-sight. We classify a cluster as aligned with the line-of-sight when it complies 
\begin{equation}
    \sum_i^M{(x_{los,i}-x_{0_{los},i})}^2>\sum_i^M{\frac{(x_{1,i}-x_{0_{1},i})^2+(x_{2,i}-x_{0_{2},i})^2}{2}}
\end{equation}
where $x_{los}$ is the line-of-sight direction, $x_{1,i}$ and $x_{2,i}$ are the directions of the projected plane and $x_{0_{los}}$, $x_{0_{1}}$, $x_{0_{2}}$ are the coordinates of the cluster center in each direction and M is the total cluster members. We use three different line-of-sight directions given by $x,y,z$ axes, and we only consider the C sample.

In Table \ref{tab:m200_aligned} we list the average of the ratios of the MAMPOSSt estimates and true values of $M_{200}$, separately for the aligned and not-aligned subsamples. As expected, MAMPOSSt tend to over-estimate the true value of $M_{200}$ for aligned halos, and to under-estimate it for not-aligned halos. When a sample of clusters is considered irrespective of the major-axis orientation with respect to the line-of-sight, a bias intermediate between the values listed in Table \ref{tab:m200_aligned} is expected, slightly closer to the vase of not-aligned clusters, since this is the most likely observing situation. This is indeed what we was found in Sect.~\ref{ss:mr3d}, Table~\ref{tab:m200}.

\begin{table}[]
\centering
\caption{MAMPOSSt vs. true $M_{200}$ ratio for aligned/not-aligned clusters (C sample)}
\label{tab:m200_aligned}
    \begin{tabular}{lcc}
    \hline
Data-set    &   \multicolumn{2}{c}{$M_{200_{MAM}}/M_{200}$} \\
            & DLB07           &     GAEA         \\ 
    \hline 
aligned         &  $1.02 \pm 0.02$  &  $1.11 \pm 0.03$   \\ 
not-aligned     &  $0.71 \pm 0.01$  &  $0.79 \pm 0.01$   \\
    \hline
    \end{tabular}
    \tablefoot{$M_{200_{MAM}}/M_{200}$ is the average of the ratios ($=3$ projections $\times$ $N_{aligned}$/$N_{not-aligned}$ clusters clusters per sample) between the MAMPOSSt determined value of $M_{200}$ and the corresponding true value. Errors are $\sigma/\sqrt{N}$.}
\end{table}

 \subsubsection{Red and blue galaxies}
 \label{sss:colors}
\begin{figure}[ht]
    \centering
    \includegraphics[height=7cm]{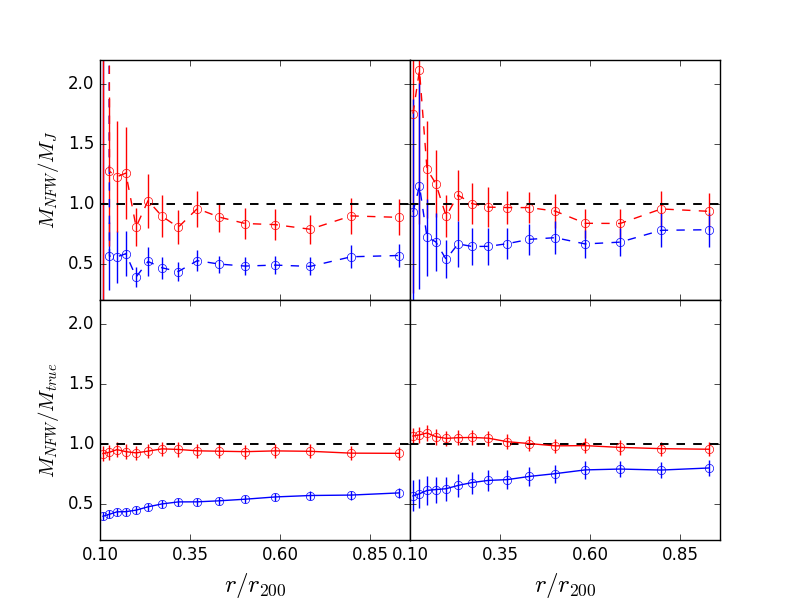}
    \caption{Ratio of different mass profile determinations obtained using red and blue galaxies as tracers, separately, for the C sample. Top panels: ratio between the median $M_{NFW}(r)$ obtained using red and blue galaxies (red and blue symbols, respectively) and $M_{Jeans}(r)$. Bottom panels: same as top panel but with $M_{true}$ in lieu of $M_{Jeans}$. Left, respectively right, panels display results obtained on the DLB07, respectively GAEA, samples.}
    \label{fig:masscolor}
\end{figure}

\begin{figure}[ht]
    \centering
    \includegraphics[height=7cm]{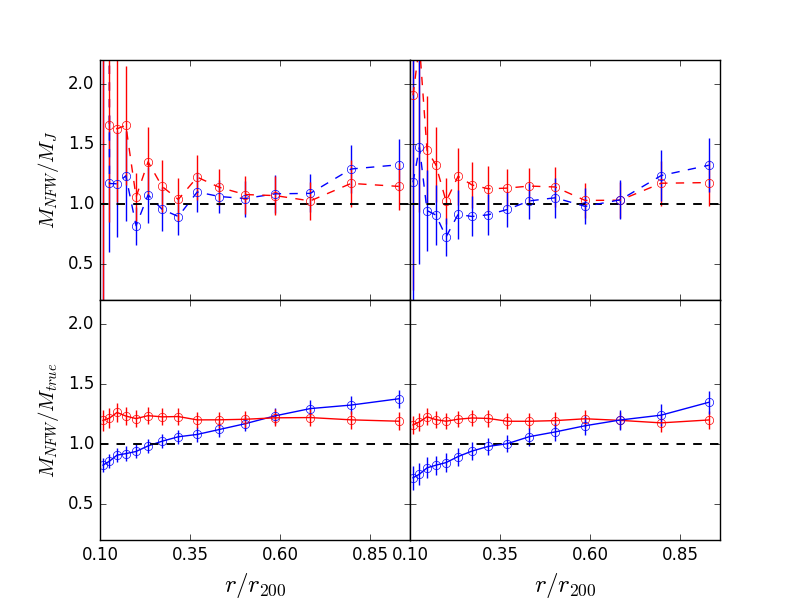}
    \caption{Ratio of different mass profile determinations obtained using red and blue galaxies as tracers, separately, for the RM sample. Symbols and colors as in Fig.~\ref{fig:masscolor}}
    \label{fig:masscolorrm}
\end{figure}

As in Sect.~\ref{sss:6dcolors} we examine here the effect of using different tracers of the gravitational potential, this time in projected phase-space. 
We compare the MAMPOSSt $M_{NFW}(r)$ obtained using red and blue galaxies as tracers, with $M_J(r)$ and $M_{true}(r)$ in Fig. \ref{fig:masscolor} for the C sample, and in Fig.~\ref{fig:masscolorrm} for the RM sample.  When using red galaxies as tracers, $M_{NFW}(r)$ is similar to $M_{true}(r)$ and $M_{Jeans}(r)$. However, when using blue galaxies as tracers, the true mass profile is strongly underestimated by MAMPOSSt at all radii. 

In Sect.~\ref{sss:6dcolors} we showed $M_{true}(r) \approx M_{Jeans}$ both for red and for blue galaxies, separately. Therefore, the failure of MAMPOSSt when using the blue population cannot be attributed to this population being outside of virial equilibrium. It cannot be attributed to the lack of blue galaxies near the cluster center either. In fact, the result for red galaxies hardly changes if we only consider red galaxies outside the inner $0.5 r_{200}$ region. The failure of MAMPOSSt is instead due to interlopers. As we commented in Sect.~\ref{sss:mr3dres}, remaining interlopers after running the Clean procedure, tend to, counter-intuitively, narrow the velocity distribution of the sample. Due to color segregation, most interlopers are blue, so the effect of interlopers is most severe for the blue galaxy population. When we repeat the analysis for the RM sample we find that $M_{NFW}/M_{J}$ is indeed similar for red and blue galaxies (see Fig.~\ref{fig:masscolorrm}).

\section{Velocity anisotropy profiles}
\label{s:beta}
In this section we present the results for the velocity anisotropy profiles $\beta(r)$ of DLB07 and GAEA clusters. We first present the true $\beta(r)$ obtained using full 6D phase-space information, then we consider how well can we reproduce these profiles using only the information from projected phase-space.
\subsection{Full phase-space}
\label{ss:beta6d}
\begin{figure}[ht]
    \centering
    \includegraphics[height=7cm]{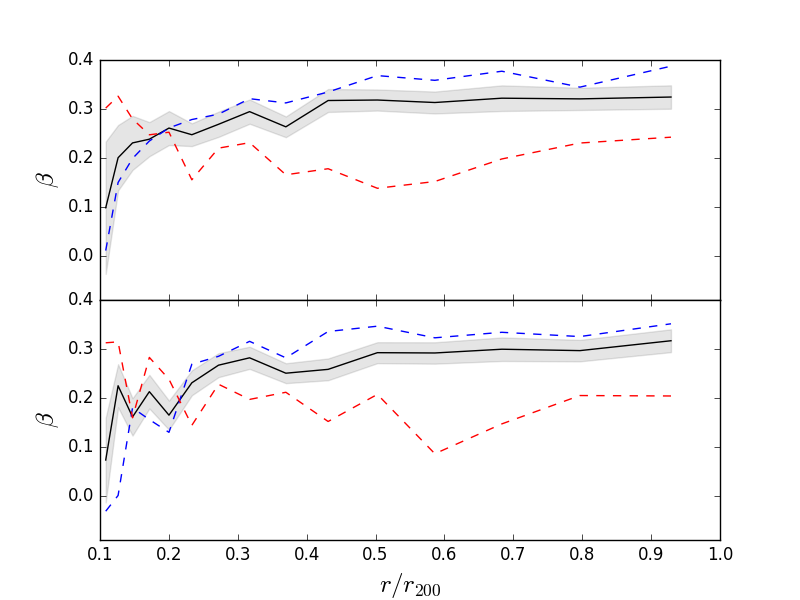}
    \caption{Median $\beta(r)$ (black solid line) for 100 clusters of the DLB07 sample (upper panel) and of the GAEA sample (lower panel). Shaded regions represent error bars $\sigma/\sqrt{N}$. 
    Blue and red dashed lines show the median $\beta(r)$ for clusters with $\chi^2$ $\leq 15$ and, respectively, $>15$, where $\chi^2$ is a value indicating the deviation of the individual cluster $\beta(r)$ from the median one.}
    \label{fig:beta6d}
\end{figure}

In Fig.~\ref{fig:beta6d} we show the median of 100 cluster $\beta(r)$, directly measured from full phase-space information, both for the DLB07 and the GAEA samples. The velocity anisotropy is close to zero near the cluster center, corresponding to isotropic orbits, and rapidly increases outside, reaching an asymptotic value of $\beta \simeq 0.3$ at $r \simeq 0.3 \, r_{200}$, corresponding to orbits with a mild radial elongation.

For each cluster we define its deviation from the median $\beta(r)$ shown in Fig.~\ref{fig:beta6d} as follows,
\begin{equation}
    \chi^2=\sum_{i=1}^{N} 
    \frac{(\beta_i-\beta_{med,i})^2}{\sigma_{i}^2},
    \label{eq:betachi2}
\end{equation}

where $\beta_{med,i}$ is the value of the median velocity anisotropy profile of the 100 galaxy clusters in the $i$-th radial bin,  $\beta_i$ is the value of velocity anisotropy profile of the given cluster in the same radial bin, $\sigma_{i}$ is its standard deviation, and $N=15$ is the number of radial bins. In the DLB07 (GAEA) sample, 59 (respectively 62) clusters have $\chi^2\leq 15$, that is their $\beta(r)$ do not deviate significantly from the median.

The median $\beta(r)$ of low- and high-$\chi^2$ $\beta(r)$ are shown in blue and red, respectively, in Fig.~\ref{fig:beta6d}. As expected, the low-$\chi^2$ median $\beta(r)$ is closer to the overall median $\beta(r)$, but in addition, there is a systematically different behaviour of the low- and high-$\chi^2$ $\beta(r)$. Low-$\chi^2$ $\beta(r)$ display an increasing radial anisotropy from the center outside, while high-$\chi^2$ $\beta(r)$ display the opposite behaviour, and a more constant $\beta$ across the $0-r_{200}$ radial range.

\begin{figure}[ht]
    \centering
    \includegraphics[height=7cm]{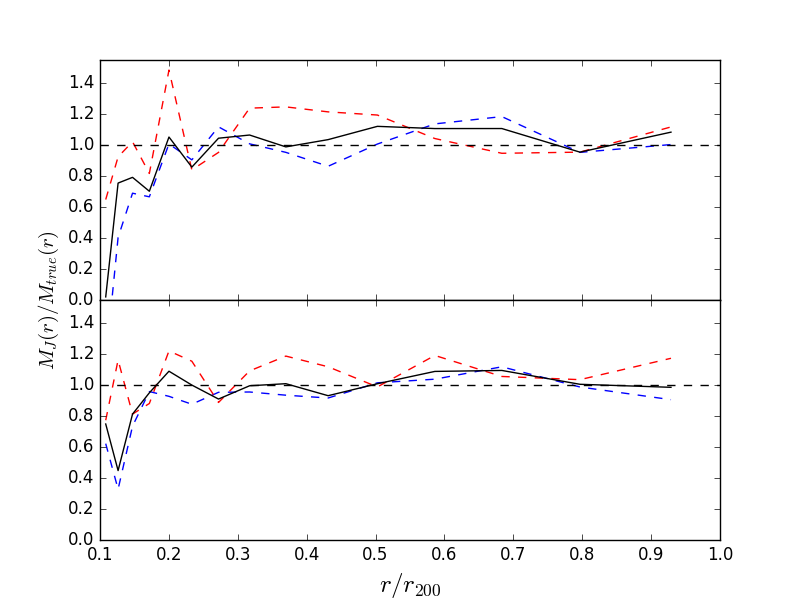}
    \caption{Median $M_J(r)/M_{true}(r)$ ratio for clusters for clusters with high-$\chi^2$ $\beta(r)$ (red-dashed line) and for clusters with low-$\chi^2$ $\beta(r)$ (blue-dashed line). The black solid line indicates the median $M_J(r)/M_{true}(r)$ for all the 100 clusters. Upper and lower panel are DLB07 and GAEA samples respectively.}
    \label{fig:massratio_chibeta}
\end{figure}

What causes some clusters to display a different $\beta(r)$ from the median? 
In Fig. \ref{fig:massratio_chibeta} we show the $M_J(r)/M_{true}(r)$ mass profile ratios for clusters with low- and high-$\chi^2$ $\beta(r)$ values, separately.  Clusters with high-$\chi^2$ $\beta(r)$ have a larger deviation from the median value of $M_J(r)/M_{true}(r)$, that is, for these clusters the Jeans equation-based mass estimate is less accurate. This partial failure of the Jeans equation could result if the cluster dynamics deviates from dynamical equilibrium, a possible consequence of cluster-cluster mergers. Deviation from dynamical equilibrium therefore seems to be accompanied with a change in the orbital distribution of cluster galaxies, from the nearly monotonic increase of $\beta$ with $r$ (median profile shown by the black curve in Fig.~\ref{fig:beta6d}) to a flatter $\beta(r)$. This change might be caused by the collision of a cluster with its infalling subclusters, generating disorder in the orbital distribution. A more detailed analysis of the orbital evolution of galaxies in simulated clusters is required to verify this scenario.

\subsubsection{Red and blue galaxies}
\label{sss:betaredblue}
Most observational studies found late-type/blue cluster galaxies to move on slightly more radial orbits than early-type/red galaxies  \citep[see, e.g.,][]{Mahdavi1999,Biviano2004,BP09,Biviano2013,MBM14,Mamon+19}, but \citet{AADDV17} found the opposite in Abell~85, and \citet{HL08} did not find any difference in the orbits of different galaxy types. Here we examine whether galaxy orbits depend on galaxy colors in the DLB07 and GAEA clusters. We adopt the red/blue color separation described in Sect.~\ref{sss:6dcolors}. 

\begin{figure}[ht]
    \centering
    \includegraphics[height=7cm]{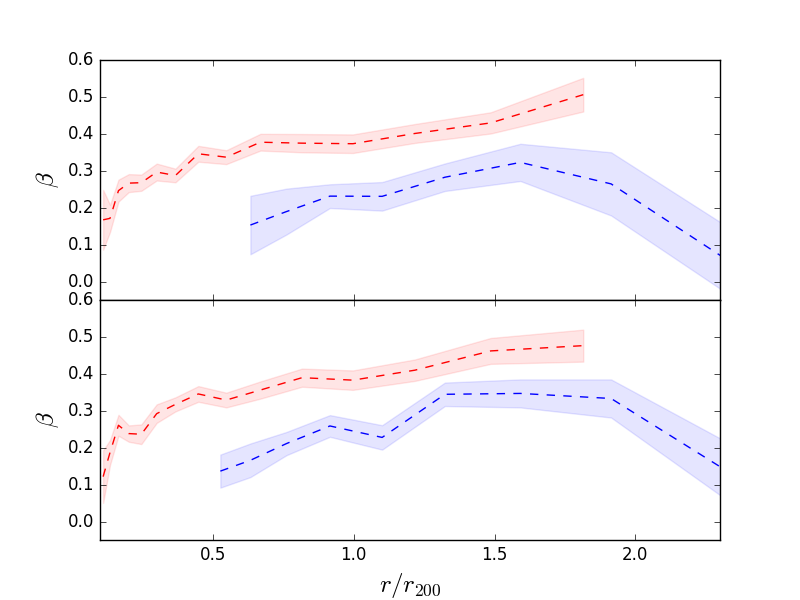}
    \caption{Median $\beta(r)$ for blue and red galaxies (blue and red dashed lines, respectively, with  $\sigma/\sqrt{N}$ confidence regions). Top panel: DLB07 sample. Bottom panel: GAEA sample}
    \label{fig:betarb}
\end{figure}

The median $\beta(r)$ of the red and blue cluster galaxy populations are different, for both the DLB07 and the GAEA sample (see
Fig.~\ref{fig:betarb}). We have shown in Sect.~\ref{sss:6dcolors} that red and blue galaxies define consistent $M_J(r)$. In fact, the
difference in the red and blue galaxies $\beta(r)$ is compensated in Jeans' eq.~(\ref{eq:jeans}) by the difference in $d \textrm{ln} \nu/d \textrm{ln} r$, red galaxies having a steeper number density profile.

Red galaxies move on more radial orbits than blue galaxies in the simulated clusters, in agreement with the results of \citet{Iannuzzi2012}, based on a SAM built on the MI, but at odds with the observational results, as well as with the results of \citet{Lotz+19}, based on hydrodynamical simulations. According to \citet{Iannuzzi2012}, galaxies on radial orbits are more subject to environmental effects and evolve from blue to red color on a shorter timescale than galaxies on more tangential (or isotropic) orbits. This would explain the prevalence of isotropic orbits among blue galaxies. On the other hand, \citet{Lotz+19} argue that we observe more radial orbits for blue galaxies because we observe them at their first passage in the cluster. After first pericentric passage these galaxies are quenched and turn to red. As these red galaxies continue orbiting the cluster they tend to be tidally disrupted at each new pericenter passage, and only galaxies on more tangential (or isotropic) orbits survive. The difference between the orbital distribution of red and blue galaxies in \citet{Iannuzzi2012} and \citet{Lotz+19} is therefore to be ascribed to the different survival time of blue galaxies in clusters in the two simulations,
longer for \citet{Iannuzzi2012} for than for \citet{Lotz+19}.

\subsubsection{Prolate and oblate clusters}
\label{sss:betashape}
\begin{figure}[ht]
    \centering
    \includegraphics[height=7cm]{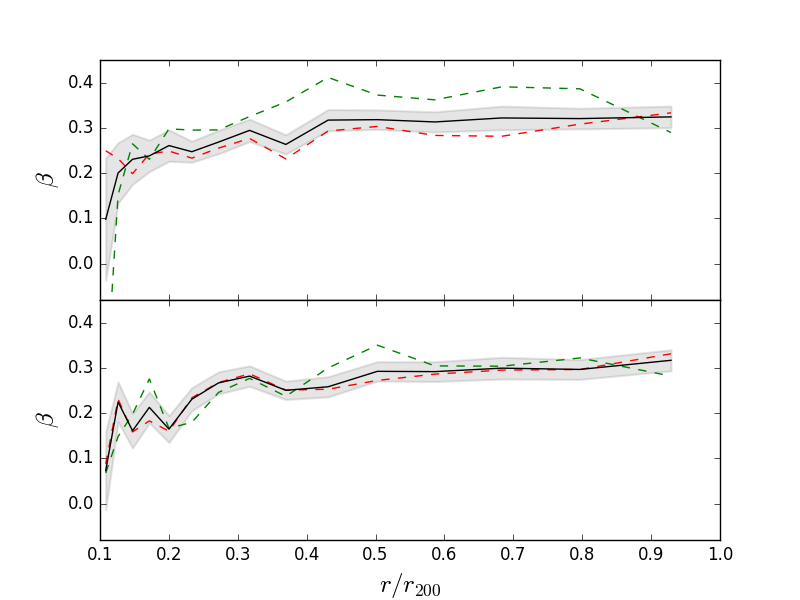}
    \caption{Median $\beta(r)$ for all clusters (solid line and $\sigma/\sqrt{N}$ uncertainty in shade), oblate clusters (dashed green lines), and prolate clusters (dashed red line). Top panel: DLB07 sample. Bottom panel: GAEA sample.}
    \label{fig:betashape}
 \end{figure}

We determine the median $\beta(r)$ separately for clusters with prolate and oblate shapes (see Sect.~\ref{sss:shapes}). In the DLB07 sample we find marginal evidence that galaxies in oblate clusters move on more radial orbits than galaxies in prolate clusters, but we find no difference in the GAEA sample (see Fig.~\ref{fig:betashape}). The difference between the two samples must be traced back to the difference in the shape distribution of clusters in DLB07 and GAEA, as seen in Fig.~\ref{fig:triaxial}, the $P$ distribution being wider in the DLB07 sample than in the GAEA sample.

The difference we observe in the DLB07 might be interpreted as a manifestation of the effect of different $\beta(r)$ along different halo axes.  In oblate clusters, one dimension is subdominant with respect the other two, while the opposite is true for prolate clusters. When sphericity is imposed, $\beta(r)$ will result from a combination of two major-axis and one minor-axis velocity anisotropy profiles  in the case of oblate clusters, and of one major-axis and two minor-axis velocity anisotropy profiles int he case of prolate clusters. Since the velocity anisotropy along the major halo axis tends to be larger than the velocity anisotropy calculated in spherical shells \citep{WGK13}, the median spherical-symmetric $\beta(r)$ of oblate clusters will tend to be larger than the corresponding quantity evaluated for prolate clusters.

\subsection{Projected phase-space}
\label{ss:betaproj}
As described in Sect.~\ref{sss:mamp},
the MAMPOSSt analysis provides the best-fit parameters not only of $M(r)$, but also of $\beta(r)$, that we model with the Tiret profile (eq.~\ref{eq:betatiret}). Here we compare the results of the MAMPOSSt analysis with the true $\beta(r)$. 

\begin{figure}[ht]
    \centering
    \includegraphics[height=7cm]{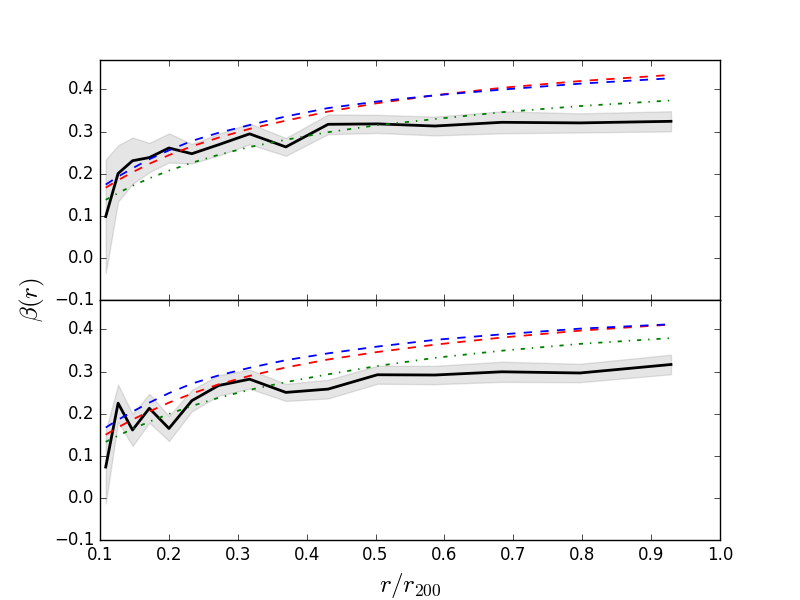}
    \caption{Median $\beta(r)$ for three orthogonal projections of 100 clusters, as obtained by the MAMPOSSt analysis, for the C (red dashed line), RM (green line), and C100 (blue line) samples.  The black solid line and shaded region indicate the $\beta(r)$ and shaded region the uncertainty $\sigma/\sqrt{N}$, obtained from full phase-space information. Upper panel: DLB07 sample. Lower panel: GAEA sample.}
    \label{fig:betamamp}
\end{figure}

In Fig.~\ref{fig:betamamp} we show the median MAMPOSSt $\beta(r)$ for the 100 clusters along three orthogonal projections, in the C (red dashed line), RM (green dot-dashed line), and C100 sample (blue dashed line). Upper and bottom panel correspond to the DLB07 and GAEA sample, respectively. The MAMPOSSt $\beta(r)$ are compared to the true median $\beta(r)$ evaluated from full phase-space information (see Sect.~\ref{ss:beta6d}). 

To quantify the comparison of the MAMPOSSt $\beta(r)$ with the true $\beta(r)$ we evaluate the $\chi^2$ statistics

\begin{equation}
\chi^2_{\beta} = \sum_{i=1}^{N} \frac{(\beta_{MAM,i}-\beta_{med,i})^2}{\sigma_{MAM,i}^2/100+\sigma_{med,i}^2},
    \label{eq:chi2betamam}
\end{equation}

where $\beta_{MAM,i}$ and $\beta_{med,i}$ are, respectively, the value of $\beta$ estimated by MAMPOSSt, and the true value of $\beta$, for the $i$-th radial bin, with $N=15$ bins, $\sigma_{med,i}$ is the error on $\beta_{med,i}$,
and $\sigma_{MAM,i}/10$ is the error on the mean of the 100 cluster values of $\beta_{MAM,i}$.
The results are given in Table~\ref{tab:beta}. The large values of $\chi^2$ indicates that the true $\beta(r)$ is not accurately reproduced. This is probably due to the uniqueness of the model adopted in the MAMPOSSt analysis, and to the fact that we allow for a single free model parameter. Moreover, the formal uncertainty in the median MAMPOSSt profile, $\sigma_{MAM,i}/10$, is rather small, $0.07$ on average. Typical uncertainties in the individual cluster MAMPOSSt $\beta_{\infty}$ values are $\sim 0.2$ \citep[see Table 2 in][]{Mamon2013a}, so that in real observations of single clusters, adopting simplified models for $\beta(r)$ is statistically acceptable.

\begin{table}[]
\centering
\caption{MAMPOSSt vs. true $\beta$}
\label{tab:beta}
    \begin{tabular}{lcc}
    \hline
Data-set    &   \multicolumn{2}{c}{$\chi^2_{\beta}$} \\
            & DLB07           &     GAEA         \\ 
    \hline 
     C    & 65 & 71 \\
     C100 & 85 & 118\\
     RM   & 21 & 37 \\
    \hline
    \end{tabular}
    \tablefoot{$\chi^2_{\beta}$ that we use to compare the MAMPOSSt determined value of $\beta$ and the corresponding true value for the different samples.}
\end{table}

The best agreement is obtained for the RM sample, even if the MAMPOSSt solution tends to over-estimate the true $\beta(r)$ at $r \gtrsim 0.5 \, r_{200}$. For the C and C100 samples, incomplete knowledge of cluster membership, that is, inclusion of interlopers among selected members, leads to a larger over-estimation of $\beta(r)$, already at $r \gtrsim 0.3 \, r_{200}$. The issue appears to be slightly more severe for the C100 than the C sample, suggesting that poor statistics might worsen the correct estimate of $\beta(r)$. 

The agreement between the true $\beta(r)$ and the MAMPOSSt solution is good at small radii, where $\beta_{med} \rightarrow 0$. However, such an agreement could be enforced by our choice of the $\beta(r)$ model, that has $\beta(0)=0$. To investigate this point, we run MAMPOSSt with four other $\beta(r)$ models,
\begin{equation}
    \beta(r) = \beta_0 + (\beta_{\infty}-\beta_0) \frac{r}{r+r_{-2}},
    \label{eq:betatmod}
\end{equation}
with $\beta_0=-5.25, -1, 0.4, 0.6$, corresponding to $\sigma_r/\sigma_{\theta}=0.4, 0.7, 1.3, 1.6$, respectively, that is, we allow for a 30 to 60\% difference in the radial and tangential components of  the velocity dispersion at the center. The Tiret model of eq.~(\ref{eq:betatiret}) is a special case of eq.~(\ref{eq:betatmod}), with $\beta_0=0$.

We adopt the Bayes Information Criterion  \citep[BIC,][]{Schwarz78} to compare the maximum likelihood values ${\cal L}$ obtained by MAMPOSSt using the Tiret model, with those obtained using the four models with non-zero central anisotropy. The median values of $\Delta {\rm BIC}=-2 \Delta \ln {\cal L}$ are listed in Table~\ref{tab:bic} for the C sample. The MAMPOSSt ${\cal L}$ are larger for the Tiret model than for $\beta_0 \neq 0$ models, and the $\Delta {\rm BIC}$ values are statistically significant \citep{KR95,Mamon+19}
for the most extreme values of central anisotropy, in particular for the GAEA sample. This analysis shows that MAMPOSSt favors $\beta_0 \approx 0$ solutions, so the agreement between the MAMPOSSt and the true $\beta(r)$ near the center is not due to the choice of the Tiret model.

\begin{table}[]
\centering
\caption{MAMPOSSt with different $\beta_0$ (C sample)}
\label{tab:bic}
    \begin{tabular}{lcc}
    \hline
$\beta_0$    &   \multicolumn{2}{c}{$\Delta {\rm BIC}$} \\
            & DLB07           &     GAEA         \\ 
    \hline 
    -5.25 & -2.0 & -5.6 \\
    -1.0  & -0.2 & -0.9 \\
     0.4  & -0.6 & -1.1 \\
     0.6  & -1.0 & -2.1 \\
     \hline
    \end{tabular}
    \tablefoot{$\Delta {\rm BIC}=-2 (\ln {\cal L}_{\rm{Tiret}} - \ln {\cal L}_{\beta_0})$, with $\beta_0$ values listed in Col.~1.}
\end{table}

\subsubsection{Line-of-sight aligned clusters}
\begin{figure}[ht]
    \centering
    \includegraphics[height=7cm]{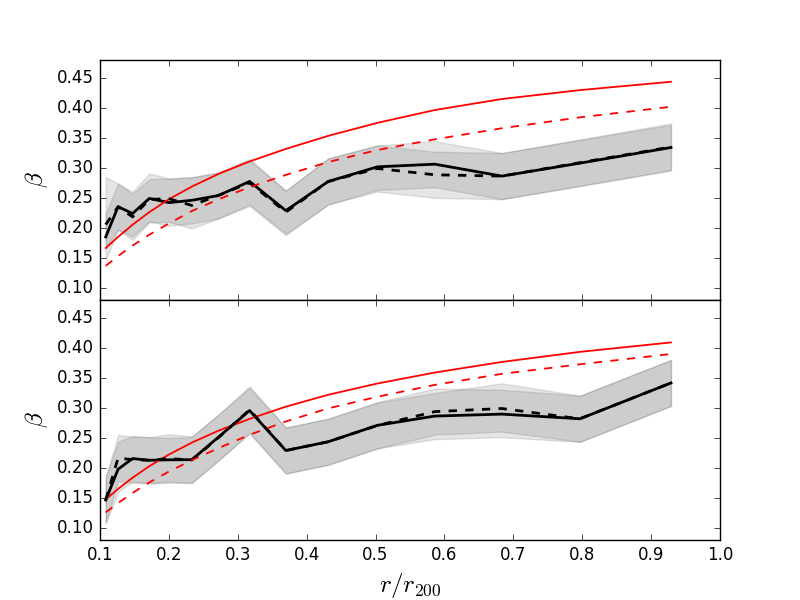}
    \caption{Median $\beta(r)$ for three lines-of-sight of aligned (dashed lines) and not-aligned (solid line) clusters for the C sample obtained by the MAMPOSSt analysis (red dashed lines).  The black lines indicate the $\beta(r)$ obtained from full phase-space information for aligned (dashed) with the of the sight and not-aligned (solid) with the of the sight, shaded region represent the uncertainty $\sigma/\sqrt{N}$. Upper panel: DLB07 sample. Lower panel: GAEA sample.}
    \label{fig:betaaligned}
\end{figure}

Using the classification of Sect. \ref{sss:shapes_aligned}, we analyze the MAMPOSSt results for $\beta(r)$ of prolate clusters with their major axis more or less aligned with the line-of-sight direction. In Fig. \ref{fig:betaaligned} we show the median $\beta(r)$ for the aligned (dashed lines) and not-aligned (solid lines) clusters of the C sample, black and red color for the true and MAMPOSSt profiles, respectively (upper and bottom panels correspond to DLB07 and GAEA samples, respectively). Aligned and not-aligned clusters have almost identical true
 $\beta(r)$ profiles as expected given that the line-of-sight only depends on the observer. The alignment along the line-of-sight has an effect on the MAMPOSSt estimates of $\beta(r)$ that are based on projected phase-space. For aligned clusters, there is a slightly better agreement of the MAMPOSSt and true $\beta(r)$ near the center, but a stronger disagreement at $r \gtrsim 0.3 \, r_{200}$, than for not-aligned clusters. More precisely, MAMPOSSt tend to over-estimate $\beta$ at large radii for both aligned and not-aligned clusters, but the over-estimate is more severe for aligned clusters.

\subsubsection{Red and blue galaxies}
\label{betamamcol}
In Sect.~\ref{sss:betaredblue} we showed that in both the DLB07 and GAEA clusters red galaxies $\beta(r)$ are larger than blue galaxies $\beta(r)$ (see Fig.~\ref{fig:betarb}), at variance with what is observed in most real clusters. Here we investigate whether this difference can be attributed to projection effects in the real clusters, by examining the MAMPOSSt solutions for the $\beta(r)$ of red and blue galaxies obtained using projected phase-space information only.

\begin{figure}[ht]
    \centering
    \includegraphics[height=7cm]{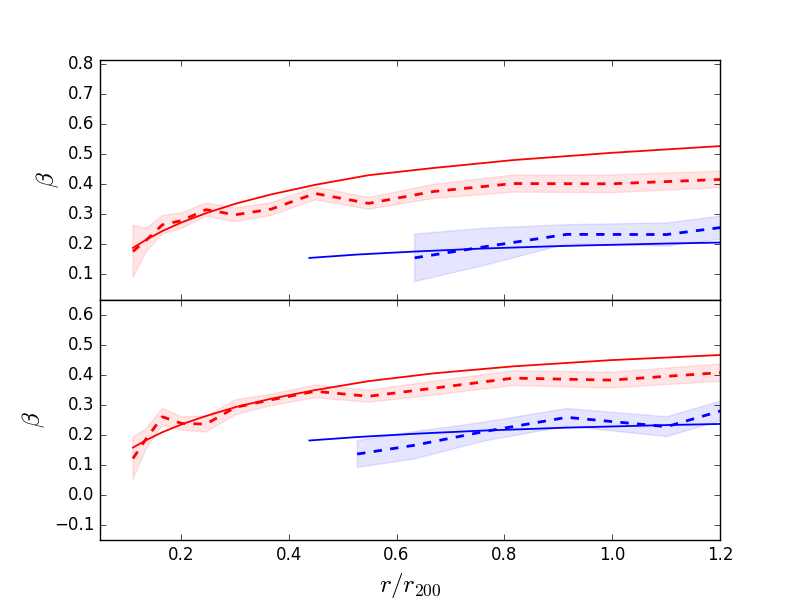}
    \caption{MAMPOSSt $\beta(r)$ estimates for red (red dashed curve and $\sigma/\sqrt{N}$ confidence level)
    and blue (blue dashed curve and $\sigma/\sqrt{N}$ confidence level) galaxies in the C sample}, compared to the true $\beta(r)$, median of 100 clusters along three orthogonal projections (solid curves with corresponding colors). Top panel: DLB07 sample. Bottom panel: GAEA sample.
    \label{fig:betamamcol}
\end{figure}

We show in Fig.~\ref{fig:betamamcol} the median $\beta(r)$ for red and blue galaxies as estimated by MAMPOSSt, for the 100
clusters along three orthogonal projections  of the DLB07 (top panel) and GAEA (lower panel) samples. In this analysis, following common practice used in observational samples, we fix the parameters of the mass profile to the best-fit values obtained using the whole cluster galaxy population. We then fit separately the number density profiles of red and blue galaxies and use the two $r_{\nu}$ best-fit parameters in two separate Split-mode runs of MAMPOSSt to determine the best-fit parameters of the red and blue galaxies $\beta(r)$.

We find that MAMPOSSt recovers the true $\beta(r)$ of both red and blue galaxies pretty well, albeit with some overestimate of the red galaxies $\beta$ at large radii. This result rules out the possibility that the discrepancy between the simulated and the observed $\beta(r)$ of red and blue galaxies (Sect.~\ref{sss:betaredblue}) is due to projection effects affecting the observational estimates in real clusters.

\section{Discussion}
\label{s:disc}

We analyzed the dynamics of two samples of 100 simulated clusters extracted from the DLB07 and GAEA data sets. The results we obtain are very similar for the two data sets. This suggests that cluster dynamics is rather insensitive to the physics of galaxy evolution. 

\subsection{Mass profiles}

When using full phase-space information we find that the Jeans equation for dynamical equilibrium of collisionless spherically symmetric systems allows an accurate estimate of the cluster mass profiles outside the central $0.2 \, r_{200}$ region. This
result is in agreement with the finding by \citet{Armmitage+18} that the velocity bias of the whole cluster population relative to the total matter is small.
We find agreement between the mass profile derived from the Jeans-equation and the true mass profile also  when using separately red and blue galaxies to trace the gravitational potential. This result implies that clusters are in dynamical equilibrium, on average, and that this is not only the case for the population of red galaxies, but also for the population of blue galaxies. The time needed for blue galaxies to reach dynamical equilibrium must be shorter than the time of color evolution in DLB07 and GAEA data-sets.

The fact that blue galaxies are on dynamical equilibrium in the cluster gravitational potential is consistent with the observational finding by \citet{Carlberg+97-equil}, who applied the Jeans equation separately to red and blue galaxies from the CNOC survey, and obtained similar mass profiles. Our result is also consistent with \citet{SMBD13} who found little difference in the mass estimates of simulated clusters when changing color selection.

At radii $r<0.2 \, r_{200}$ we find that application of the Jeans equation leads to an under-estimate of the true mass profile. We attribute this under-estimate to the effect of dynamical friction, a process that can invalidate the assumption of collisionless tracers of the Jeans equation. 
It is not the purpose of this paper to test how closely the dynamical friction process implemented in the DLB07 and GAEA simulations resembles the physical process at work in real clusters of galaxies. In any case, other simulations have found evidence that dynamical friction affects the dynamics of massive cluster galaxies \citep[e.g.][]{SWTK01}. \citet{SMBD13} found that dynamical friction can introduce a significant velocity bias when only the 30 brightest galaxies of a simulated cluster are selected as a tracer of the gravitational potential. Evidence of this same effect has been found in observations of real clusters \citep{BGMM92,Old+13}. 

While the collisionless assumption appears to be a problem for the application of the Jeans equation, the assumption of spherical symmetry is not, as we checked by considering prolate and oblate clusters separately. Our conclusion is in agreement with previous results \citep{PJS03,SMBD13}. 

In projected phase space, the combination of the Clean selection of cluster members and MAMPOSSt dynamical analysis provides slightly biased result for the mass profile, with an underestimate of the true mass profile by 7-17\% and a standard deviation of 35-69\%. These results are not as good as those indicated by \citet{Mamon2013a}. We also characterize the mass bias as a function of radius, and find it to be nearly constant at all radii. We attribute the negative mass bias to the presence of interlopers, as initially suggested by \citet{Cen1997}.
Our conclusion is supported by the fact that considering only red galaxies, the mass bias is less significant. This has been already suggested by other studies \citep{Biviano2006,MBM10}, but here we have determined the mass bias as a function of radius, for the first time for an estimator based on projected phase-space distribution of galaxies \citep[see][for the mass profile bias from lensing and X-ray data]{Rasia+12}, and found this mass bias to be independent of radius. 

While we found that the effect of cluster asphericity is not important, deviation from spherical symmetry can be important for individual clusters when observed in projection, if, for instance, the line-of-sight is aligned with the main axis of a prolate cluster \citep{Wojtak2013}. Indeed we find that the bias in the cluster mass estimate depends on the orientation of the cluster major axis with respect to the line-of-sight. The mass of aligned clusters is over-estimated in projection, while that of not-aligned clusters is under-estimated.

\subsection{Velocity anisotropy profiles}
The velocity anisotropy profiles of simulated clusters from the DLB07 and GAEA data-sets are very similar, slightly radial at all radii, increasing from the center outside and reaching a plateau of $\beta \sim 0.3$ at $r \gtrsim 0.3 r_{200}$, in substantial agreement with several previous investigations of both simulated \citep{Diaferio99,MBM10,Mamon2013a,Munari2013} and real \citep{NK96,Lemze+09,Biviano2013,Capasso+19} clusters. We do not find a significant difference in the $\beta(r)$ of oblate and prolate clusters, so we argue that cluster asphericity is not a major problem for the determination of cluster galaxies orbits using the spherical Jeans equation.

We find that clusters for which the Jeans equation provides a less accurate estimate of the true mass profile have a flatter velocity anisotropy profile. Galaxies in clusters that are less dynamically relaxed move on more radial orbits near the cluster center, and more isotropic orbits in the cluster outskirts, compared to galaxies in dynamically relaxed clusters.

Clusters are expected to reach dynamical equilibrium through a phase of violent relaxation that leads to erasure of any velocity anisotropy acquired during the initial collapse phase \citep{LyndenBell67}. The cluster then grows in mass during the subsequent phase of smooth accretion, that preserves the orbital anisotropy of the infalling material \citep{LC11}. The resulting $\beta(r) \approx 0$ (isotropic orbits) near the center, and $\beta(r) >0$ (radial orbits) outside \citep[see Fig.~14 in][]{LC11}, is similar to the profile we find for dynamically relaxed clusters (blue dashed line in Fig.~\ref{fig:beta6d}). When a major merger occur between the cluster and an accreting subcluster, chaotic mixing \citep{KS03} is likely to erase any pre-existing order in the radial orbital distribution. This could lead to a flattening of $\beta(r)$, as observed in our dynamically unrelaxed clusters (red dashed line in Fig.~\ref{fig:beta6d}). We plan to test this scenario in the future, by analysing the evolution of the orbital distribution of galaxies in simulated halos.

When we examine the orbits of red and blue galaxies in simulated clusters from DLB07 and GAEA we find  a different result than in the observational studies of \citep[e.g.,][]{Mahdavi1999,Biviano2004,BP09,Biviano2013,MBM14,Mamon+19}, although in agreement with  \citet{AADDV17} for the cluster Abell~85. Red galaxies have more radial orbits than the blue ones, consistent with the results obtained in the simulations by \citet{Iannuzzi2012}. According to these authors the different orbits can be explained by the environmentally-driven color evolution of cluster galaxies. Environmental effects are expected to be stronger close to the cluster center, so that galaxies on more radial orbits, with a smaller pericentric radius, are more strongly affected by environmental effects than galaxies on more isotropic orbits. Only galaxies that avoid the central cluster regions, by moving on less radial orbits, can avoid the fate of becoming red. \citet{Iannuzzi2012} do not consider orphan galaxies in their analysis, while we do. If we exclude orphans from our analysis, the results on the orbits of red and blue galaxies remain consistent with \citet{Iannuzzi2012}.  

Using hydrodynamical simulations, \citet{Lotz+19} find a more radial orbital distribution of blue galaxies than we find, in better agreement with observations. These authors argue that we only observe blue galaxies on their first infall, as they are quenched at first pericentric passage and turn red, a scenario supported by observations \citep{Mamon+19}. Subsequent pericentric passages can destroy these red galaxies, the more so the more radial are their orbits. As a consequence, we only observe radial orbits for galaxies on first infall, when they are blue. Red galaxies that survive are those on more isotropic orbits.

In projected phase-space MAMPOSSt is able to recover the true velocity anisotropy profiles with a slight over-estimate at $r \gtrsim 0.3 \, r_{200}$ when the C and C100 samples are considered. This over-estimate can be attributed to the presence of interlopers. The agreement is somewhat better for the RM sample, and somewhat worse for prolate clusters with their major axis aligned with the line-of-sight direction.

MAMPOSSt is also capable of reproducing the true velocity anisotropy profiles of red and blue galaxies separately. Note that this result is obtained by first determining the mass profile with MAMPOSSt using all galaxies, and then by using this mass profile determination to find the velocity anisotropy profiles of red and blue galaxies, separately. Given that MAMPOSSt can correctly reproduce the velocity anisotropy profiles of two galaxy populations separately, the mismatch between the simulated velocity anisotropy profiles of red and blue galaxies and several observational studies is not likely due to a problem in the analysis of observational sample. This mismatch suggests that some evolutionary process of cluster galaxies is not correctly implemented in the simulations we investigate. However, what is more relevant here is not how close are the simulated clusters to the real ones, but how close are the MAMPOSSt profiles, based on projected phase-space data, to the simulated ones.

\section{Conclusions}\label{s:conc}
Using the DLB07 and GAEA data sets of 100 galaxy clusters from the MI, we investigate the mass and the velocity anisotropy profiles of galaxy clusters, using the spatial and velocity distributions of cluster galaxies both in full and projected phase-space. The results obtained for the two data sets are similar.

For the mass profiles, considering full phase-space information, we find that outside the central $0.2 \, r_{200}$ region, there is an agreement between the mass profile obtained from the Jeans equation for dynamical equilibrium of collisionless spherically symmetric systems and the real mass profiles. This indicates that the simulated clusters have reached dynamical equilibrium. This agreement remains also for mass profiles obtained from the Jeans-equation using red and blue galaxies separately as tracers, indicating that both populations of galaxies are in dynamical equilibrium in the cluster potential. For the central regions, $r<0.2 \, r_{200}$, the profile obtained from the Jeans equation underestimates the true mass profile, and we attribute this discrepancy to dynamical friction invalidating the collisionless fluid assumption in the Jeans equation. On the other hand, the spherical assumption of does not seem to be a problem, despite cluster asphericity, neither for prolate, nor for oblate clusters.

In projected phase-space, we simulate observational procedures, by combining the Clean algorithm of members selection and the MAMPOSSt dynamical analysis. We find a bias in the derived mass profiles of 7-17\%, similar at all radii, except very near the center. This bias does not depend on the number of cluster members, when at least 100 members are considered.
We argue that the bias is due to the imperfect removal of interlopers, and we find that the bias is reduced when we only consider red galaxies as tracers of the gravitational potential, and  it disappears when considering true cluster members.

The velocity anisotropy profiles are slightly radial and increase from the center outside, reaching a plateau $\beta \sim 0.3$ at $r \gtrsim 0.3 r_{200}$. The anisotropy profile of less dynamically relaxed clusters is flatter. We argue that this could be the effect of orbital re-distribution following cluster-subcluster major mergers. Red galaxies move on more radially elongated orbits than blue galaxies, at variance with what is found in most real clusters. 

Using projected phase-space information, and adopting a rather simple model for $\beta(r)$, MAMPOSSt estimates the true velocity anisotropy profiles rather accurately, albeit with a slight over-estimate of the radial anisotropy at radii $r \gtrsim 0.3 \, r_{200}$. Such over-estimate is more severe for prolate clusters observed with their major axis aligned along the line-of-sight. The MAMPOSSt estimates are similarly accurate for the $\beta(r)$ of red and blue galaxies considered separately. This indicates that the $\beta(r)$ estimated for red and blue galaxies in real clusters are not biased. The fact that they are different from the intrinsic $\beta(r)$ of DLB07 and GAEA simulated clusters then suggests that some evolutionary processes of galaxies in clusters are not correctly implemented in the simulations.

Our study indicates that state of the art modelling of the internal dynamics of clusters is a robust and viable tool to determine the cluster mass and velocity anisotropy profiles. Future studies with different modelling of galaxy astrophysics, including comparison to observations, may be used to deepen our understanding of the interplay between galaxy orbits and their evolution in clusters, as well as to develop more accurate and precise methods of mass profile determinations.

\begin{acknowledgements}
      We thank the referee for her/his useful comments that contributed to the scientific content of this paper. This work has been largely supported by the LACEGAL program. AB thanks IATE for hospitality. 
\end{acknowledgements}

\bibliography{references}

\begin{thebibliography}{73}
\expandafter\ifx\csname natexlab\endcsname\relax\def\natexlab#1{#1}\fi

\bibitem[{{Adami} {et~al.}(1998){Adami}, {Mazure}, {Katgert}, \&
  {Biviano}}]{AMKB98}
{Adami}, C., {Mazure}, A., {Katgert}, P., \& {Biviano}, A. 1998, \aap, 336, 63

\bibitem[{{Aguerri} {et~al.}(2017){Aguerri}, {Agulli}, {Diaferio}, \& {Dalla
  Vecchia}}]{AADDV17}
{Aguerri}, J.~A.~L., {Agulli}, I., {Diaferio}, A., \& {Dalla Vecchia}, C. 2017,
  \mnras, 468, 364

\bibitem[{{Allen}(1998)}]{Allen98}
{Allen}, S.~W. 1998, \mnras, 296, 392

\bibitem[{{Armitage} {et~al.}(2018){Armitage}, {Barnes}, {Kay}, {Bah{\'e}},
  {Dalla Vecchia}, {Crain}, \& {Theuns}}]{Armmitage+18}
{Armitage}, T.~J., {Barnes}, D.~J., {Kay}, S.~T., {et~al.} 2018, \mnras, 474,
  3746

\bibitem[{{Bartelmann}(1996)}]{Bartelmann1996}
{Bartelmann}, M. 1996, \aap, 313, 697

\bibitem[{{Binney} \& {Tremaine}(1987)}]{Binney1987}
{Binney}, J. \& {Tremaine}, S. 1987, {Galactic dynamics}

\bibitem[{{Biviano}(2020)}]{Biviano20}
{Biviano}, A. 2020, Boletin de la Asociacion Argentina de Astronomia La Plata
  Argentina, 61B, 142, arXiv:2001.00800

\bibitem[{{Biviano} {et~al.}(1992){Biviano}, {Girardi}, {Giuricin},
  {Mardirossian}, \& {Mezzetti}}]{BGMM92}
{Biviano}, A., {Girardi}, M., {Giuricin}, G., {Mardirossian}, F., \&
  {Mezzetti}, M. 1992, \apj, 396, 35

\bibitem[{{Biviano} \& {Katgert}(2004)}]{Biviano2004}
{Biviano}, A. \& {Katgert}, P. 2004, \aap, 424, 779

\bibitem[{{Biviano} {et~al.}(2017){Biviano}, {Moretti}, {Paccagnella},
  {Poggianti}, {Bettoni}, {Gullieuszik}, {Vulcani}, {Fasano}, {D'Onofrio},
  {Fritz}, \& {Cava}}]{Biviano+17a}
{Biviano}, A., {Moretti}, A., {Paccagnella}, A., {et~al.} 2017, \aap, 607, A81

\bibitem[{{Biviano} {et~al.}(2006){Biviano}, {Murante}, {Borgani}, {Diaferio},
  {Dolag}, \& {Girardi}}]{Biviano2006}
{Biviano}, A., {Murante}, G., {Borgani}, S., {et~al.} 2006, \aap, 456, 23

\bibitem[{{Biviano} \& {Poggianti}(2009)}]{BP09}
{Biviano}, A. \& {Poggianti}, B.~M. 2009, \aap, 501, 419

\bibitem[{{Biviano} {et~al.}(2013){Biviano}, {Rosati}, {Balestra}, {Mercurio},
  {Girardi}, {Nonino}, {Grillo}, {Scodeggio}, {Lemze}, \&
  {Kelson}}]{Biviano2013}
{Biviano}, A., {Rosati}, P., {Balestra}, I., {et~al.} 2013, \aap, 558, A1

\bibitem[{{Biviano} {et~al.}(2016){Biviano}, {van der Burg}, {Muzzin},
  {Sartoris}, {Wilson}, \& {Yee}}]{Biviano2016}
{Biviano}, A., {van der Burg}, R.~F.~J., {Muzzin}, A., {et~al.} 2016, \aap,
  594, A51

\bibitem[{{Capasso} {et~al.}(2019){Capasso}, {Saro}, {Mohr}, {Biviano},
  {Bocquet}, {Strazzullo}, {Grandis}, {Applegate}, {Bayliss}, {Benson},
  {Bleem}, {Brodwin}, {Bulbul}, {Carlstrom}, {Chiu}, {Dietrich}, {Gupta}, {de
  Haan}, {Hlavacek-Larrondo}, {Klein}, {von der Linden}, {McDonald}, {Rapetti},
  {Reichardt}, {Sharon}, {Stalder}, {Stanford}, {Stark}, {Stern}, \&
  {Zenteno}}]{Capasso+19}
{Capasso}, R., {Saro}, A., {Mohr}, J.~J., {et~al.} 2019, \mnras, 482, 1043

\bibitem[{{Carlberg} {et~al.}(1997){Carlberg}, {Yee}, {Ellingson}, {Morris},
  {Abraham}, {Gravel}, {Pritchet}, {Smecker-Hane}, {Hartwick}, {Hesser},
  {Hutchings}, \& {Oke}}]{Carlberg+97-equil}
{Carlberg}, R.~G., {Yee}, H. K.~C., {Ellingson}, E., {et~al.} 1997, \apjl, 476,
  L7

\bibitem[{{Cen}(1997)}]{Cen1997}
{Cen}, R. 1997, \apj, 485, 39

\bibitem[{{Clowe} {et~al.}(2006){Clowe}, {Brada{\v c}}, {Gonzalez},
  {Markevitch}, {Randall}, {Jones}, \& {Zaritsky}}]{Clowe+06b}
{Clowe}, D., {Brada{\v c}}, M., {Gonzalez}, A.~H., {et~al.} 2006, \apjl, 648,
  L109

\bibitem[{{Cole}(1991)}]{Cole1991}
{Cole}, S. 1991, \apj, 367, 45

\bibitem[{{De Lucia} \& {Blaizot}(2007)}]{DeLucia2007}
{De Lucia}, G. \& {Blaizot}, J. 2007, \mnras, 375, 2

\bibitem[{{De Lucia} {et~al.}(2019){De Lucia}, {Hirschmann}, \&
  {Fontanot}}]{DeLucia2019}
{De Lucia}, G., {Hirschmann}, M., \& {Fontanot}, F. 2019, \mnras, 482, 5041

\bibitem[{{De Lucia} {et~al.}(2014){De Lucia}, {Tornatore}, {Frenk}, {Helmi},
  {Navarro}, \& {White}}]{DeLucia2014}
{De Lucia}, G., {Tornatore}, L., {Frenk}, C.~S., {et~al.} 2014, \mnras, 445,
  970

\bibitem[{{Diaferio}(1999)}]{Diaferio99}
{Diaferio}, A. 1999, \mnras, 309, 610

\bibitem[{{Diemer} \& {Kravtsov}(2014)}]{DK14}
{Diemer}, B. \& {Kravtsov}, A.~V. 2014, \apj, 789, 1

\bibitem[{{Ettori} {et~al.}(2002){Ettori}, {De Grandi}, \& {Molendi}}]{EDGM02}
{Ettori}, S., {De Grandi}, S., \& {Molendi}, S. 2002, \aap, 391, 841

\bibitem[{{Evrard} {et~al.}(2008){Evrard}, {Bialek}, {Busha}, {White}, {Habib},
  {Heitmann}, {Warren}, {Rasia}, {Tormen}, {Moscardini}, {Power}, {Jenkins},
  {Gao}, {Frenk}, {Springel}, {White}, \& {Diemand}}]{Evrard+08}
{Evrard}, A.~E., {Bialek}, J., {Busha}, M., {et~al.} 2008, \apj, 672, 122

\bibitem[{{Fontanot} {et~al.}(2009){Fontanot}, {Somerville}, {Silva}, {Monaco},
  \& {Skibba}}]{Fontanot2009}
{Fontanot}, F., {Somerville}, R.~S., {Silva}, L., {Monaco}, P., \& {Skibba}, R.
  2009, \mnras, 392, 553

\bibitem[{{Hirschmann} {et~al.}(2016){Hirschmann}, {De Lucia}, \&
  {Fontanot}}]{Hirsch2016}
{Hirschmann}, M., {De Lucia}, G., \& {Fontanot}, F. 2016, \mnras, 461, 1760

\bibitem[{{Hoekstra} {et~al.}(2004){Hoekstra}, {Yee}, \& {Gladders}}]{HYG04}
{Hoekstra}, H., {Yee}, H.~K.~C., \& {Gladders}, M.~D. 2004, \apj, 606, 67

\bibitem[{{Hwang} \& {Lee}(2008)}]{HL08}
{Hwang}, H.~S. \& {Lee}, M.~G. 2008, \apj, 676, 218

\bibitem[{{Iannuzzi} \& {Dolag}(2012)}]{Iannuzzi2012}
{Iannuzzi}, F. \& {Dolag}, K. 2012, \mnras, 427, 1024

\bibitem[{{Joshi} {et~al.}(2020){Joshi}, {Pillepich}, {Nelson}, {Marinacci},
  {Springel}, {Rodriguez-Gomez}, {Vogelsberger}, \& {Hernquist}}]{Joshi+20}
{Joshi}, G.~D., {Pillepich}, A., {Nelson}, D., {et~al.} 2020, \mnras, 496, 2673

\bibitem[{{Kandrup} \& {Siopis}(2003)}]{KS03}
{Kandrup}, H.~E. \& {Siopis}, C. 2003, \mnras, 345, 727

\bibitem[{{Kass} \& {Rafferty}(1995)}]{KR95}
{Kass}, R. \& {Rafferty}, A. 1995, J. Am. Stat. Assoc., 90, 773

\bibitem[{{Kasun} \& {Evrard}(2005)}]{Kasun2005}
{Kasun}, S.~F. \& {Evrard}, A.~E. 2005, \apj, 629, 781

\bibitem[{{Lapi} \& {Cavaliere}(2011)}]{LC11}
{Lapi}, A. \& {Cavaliere}, A. 2011, \apj, 743, 127

\bibitem[{{Lemze} {et~al.}(2009){Lemze}, {Broadhurst}, {Rephaeli}, {Barkana},
  \& {Umetsu}}]{Lemze+09}
{Lemze}, D., {Broadhurst}, T., {Rephaeli}, Y., {Barkana}, R., \& {Umetsu}, K.
  2009, \apj, 701, 1336

\bibitem[{{Limousin} {et~al.}(2013){Limousin}, {Morandi}, {Sereno},
  {Meneghetti}, {Ettori}, {Bartelmann}, \& {Verdugo}}]{Limousin2013}
{Limousin}, M., {Morandi}, A., {Sereno}, M., {et~al.} 2013, \ssr, 177, 155

\bibitem[{{Lotz} {et~al.}(2019){Lotz}, {Remus}, {Dolag}, {Biviano}, \&
  {Burkert}}]{Lotz+19}
{Lotz}, M., {Remus}, R.-S., {Dolag}, K., {Biviano}, A., \& {Burkert}, A. 2019,
  \mnras, 488, 5370

\bibitem[{{Lynden-Bell}(1967)}]{LyndenBell67}
{Lynden-Bell}, D. 1967, \mnras, 136, 101

\bibitem[{{Macci{\`o}} {et~al.}(2008){Macci{\`o}}, {Dutton}, \& {van den
  Bosch}}]{Maccio2008}
{Macci{\`o}}, A.~V., {Dutton}, A.~A., \& {van den Bosch}, F.~C. 2008, \mnras,
  391, 1940

\bibitem[{{Mahdavi} {et~al.}(1999){Mahdavi}, {Geller}, {B{\"o}hringer},
  {Kurtz}, \& {Ramella}}]{Mahdavi1999}
{Mahdavi}, A., {Geller}, M.~J., {B{\"o}hringer}, H., {Kurtz}, M.~J., \&
  {Ramella}, M. 1999, \apj, 518, 69

\bibitem[{{Mamon} {et~al.}(2013){Mamon}, {Biviano}, \& {Bou{\'e}}}]{Mamon2013a}
{Mamon}, G.~A., {Biviano}, A., \& {Bou{\'e}}, G. 2013, \mnras, 429, 3079

\bibitem[{{Mamon} {et~al.}(2010){Mamon}, {Biviano}, \& {Murante}}]{MBM10}
{Mamon}, G.~A., {Biviano}, A., \& {Murante}, G. 2010, \aap, 520, A30

\bibitem[{{Mamon} \& {Bou{\'e}}(2010)}]{Mamon2010}
{Mamon}, G.~A. \& {Bou{\'e}}, G. 2010, \mnras, 401, 2433

\bibitem[{{Mamon} {et~al.}(2019){Mamon}, {Cava}, {Biviano}, {Moretti},
  {Poggianti}, \& {Bettoni}}]{Mamon+19}
{Mamon}, G.~A., {Cava}, A., {Biviano}, A., {et~al.} 2019, \aap, 631, A131

\bibitem[{{Meneghetti} {et~al.}(2014){Meneghetti}, {Rasia}, {Vega}, {Merten},
  {Postman}, {Yepes}, {Sembolini}, {Donahue}, {Ettori}, {Umetsu}, {Balestra},
  {Bartelmann}, {Ben{\'{\i}}tez}, {Biviano}, {Bouwens}, {Bradley},
  {Broadhurst}, {Coe}, {Czakon}, {De Petris}, {Ford}, {Giocoli},
  {Gottl{\"o}ber}, {Grillo}, {Infante}, {Jouvel}, {Kelson}, {Koekemoer},
  {Lahav}, {Lemze}, {Medezinski}, {Melchior}, {Mercurio}, {Molino},
  {Moscardini}, {Monna}, {Moustakas}, {Moustakas}, {Nonino}, {Rhodes},
  {Rosati}, {Sayers}, {Seitz}, {Zheng}, \& {Zitrin}}]{Meneghetti+14}
{Meneghetti}, M., {Rasia}, E., {Vega}, J., {et~al.} 2014, \apj, 797, 34

\bibitem[{{Merritt}(1987)}]{Merritt87}
{Merritt}, D. 1987, \apj, 313, 121

\bibitem[{{Munari} {et~al.}(2013){Munari}, {Biviano}, {Borgani}, {Murante}, \&
  {Fabjan}}]{Munari2013}
{Munari}, E., {Biviano}, A., {Borgani}, S., {Murante}, G., \& {Fabjan}, D.
  2013, \mnras, 430, 2638

\bibitem[{{Munari} {et~al.}(2014){Munari}, {Biviano}, \& {Mamon}}]{MBM14}
{Munari}, E., {Biviano}, A., \& {Mamon}, G.~A. 2014, \aap, 566, A68

\bibitem[{{Natarajan} \& {Kneib}(1996)}]{NK96}
{Natarajan}, P. \& {Kneib}, J.-P. 1996, \mnras, 283, 1031

\bibitem[{{Navarro} {et~al.}(1997){Navarro}, {Frenk}, \& {White}}]{Navarro1997}
{Navarro}, J.~F., {Frenk}, C.~S., \& {White}, S. D.~M. 1997, \apj, 490, 493

\bibitem[{{Old} {et~al.}(2013){Old}, {Gray}, \& {Pearce}}]{Old+13}
{Old}, L., {Gray}, M.~E., \& {Pearce}, F.~R. 2013, \mnras, 434, 2606

\bibitem[{{Old} {et~al.}(2015){Old}, {Wojtak}, {Mamon}, {Skibba}, {Pearce},
  {Croton}, {Bamford}, {Behroozi}, {de Carvalho}, {Mu{\~n}oz-Cuartas},
  {Gifford}, {Gray}, {der Linden}, {Merrifield}, {Muldrew}, {M{\"u}ller},
  {Pearson}, {Ponman}, {Rozo}, {Rykoff}, {Saro}, {Sepp}, {Sif{\'o}n}, \&
  {Tempel}}]{Old+15}
{Old}, L., {Wojtak}, R., {Mamon}, G.~A., {et~al.} 2015, \mnras, 449, 1897

\bibitem[{{Old} {et~al.}(2018){Old}, {Wojtak}, {Pearce}, {Gray}, {Mamon},
  {Sif{\'o}n}, {Tempel}, {Biviano}, {Yee}, {de Carvalho}, {M{\"u}ller}, {Sepp},
  {Skibba}, {Croton}, {Bamford}, {Power}, {von der Linden}, \& {Saro}}]{Old+18}
{Old}, L., {Wojtak}, R., {Pearce}, F.~R., {et~al.} 2018, \mnras, 475, 853

\bibitem[{{Paz} {et~al.}(2006){Paz}, {Lambas}, {Mercjam}, \&
  {Padilla}}]{Paz2006}
{Paz}, D.~J., {Lambas}, D.~G., {Mercjam}, M., \& {Padilla}, N.~D. 2006, in
  Revista Mexicana de Astronomia y Astrofisica Conference Series, Vol.~26, 195

\bibitem[{{Piffaretti} {et~al.}(2003){Piffaretti}, {Jetzer}, \&
  {Schindler}}]{PJS03}
{Piffaretti}, R., {Jetzer}, P., \& {Schindler}, S. 2003, \aap, 398, 41

\bibitem[{{Pratt} {et~al.}(2019){Pratt}, {Arnaud}, {Biviano}, {Eckert},
  {Ettori}, {Nagai}, {Okabe}, \& {Reiprich}}]{Pratt+19}
{Pratt}, G.~W., {Arnaud}, M., {Biviano}, A., {et~al.} 2019, \ssr, 215, 25

\bibitem[{{Rasia} {et~al.}(2013){Rasia}, {Borgani}, {Ettori}, {Mazzotta}, \&
  {Meneghetti}}]{Rasia+13}
{Rasia}, E., {Borgani}, S., {Ettori}, S., {Mazzotta}, P., \& {Meneghetti}, M.
  2013, \apj, 776, 39

\bibitem[{{Rasia} {et~al.}(2012){Rasia}, {Meneghetti}, {Martino}, {Borgani},
  {Bonafede}, {Dolag}, {Ettori}, {Fabjan}, {Giocoli}, {Mazzotta}, {Merten},
  {Radovich}, \& {Tornatore}}]{Rasia+12}
{Rasia}, E., {Meneghetti}, M., {Martino}, R., {et~al.} 2012, New Journal of
  Physics, 14, 055018

\bibitem[{{Saro} {et~al.}(2013){Saro}, {Mohr}, {Bazin}, \& {Dolag}}]{SMBD13}
{Saro}, A., {Mohr}, J.~J., {Bazin}, G., \& {Dolag}, K. 2013, \apj, 772, 47

\bibitem[{{Schwarz}(1978)}]{Schwarz78}
{Schwarz}, G. 1978, Ann. Stat., 6, 461

\bibitem[{{Springel} {et~al.}(2005){Springel}, {White}, {Jenkins}, {Frenk},
  {Yoshida}, {Gao}, {Navarro}, {Thacker}, {Croton}, {Helly}, {Peacock}, {Cole},
  {Thomas}, {Couchman}, {Evrard}, {Colberg}, \& {Pearce}}]{Springel2005}
{Springel}, V., {White}, S. D.~M., {Jenkins}, A., {et~al.} 2005, \nat, 435, 629

\bibitem[{{Springel} {et~al.}(2001){Springel}, {White}, {Tormen}, \&
  {Kauffmann}}]{SWTK01}
{Springel}, V., {White}, S. D.~M., {Tormen}, G., \& {Kauffmann}, G. 2001,
  \mnras, 328, 726

\bibitem[{{Tiret} {et~al.}(2007){Tiret}, {Combes}, {Angus}, {Famaey}, \&
  {Zhao}}]{Tiret+07}
{Tiret}, O., {Combes}, F., {Angus}, G.~W., {Famaey}, B., \& {Zhao}, H.~S. 2007,
  \aap, 476, L1

\bibitem[{{Tonnesen}(2019)}]{Tonnesen19}
{Tonnesen}, S. 2019, \apj, 874, 161

\bibitem[{{van der Marel} {et~al.}(2000){van der Marel}, {Magorrian},
  {Carlberg}, {Yee}, \& {Ellingson}}]{vanderMarel+00}
{van der Marel}, R.~P., {Magorrian}, J., {Carlberg}, R.~G., {Yee}, H.~K.~C., \&
  {Ellingson}, E. 2000, \aj, 119, 2038

\bibitem[{{White} \& {Frenk}(1991)}]{White&Frenk1991}
{White}, S. D.~M. \& {Frenk}, C.~S. 1991, \apj, 379, 52

\bibitem[{{Wojtak}(2013)}]{Wojtak2013}
{Wojtak}, R. 2013, \aap, 559, A89

\bibitem[{{Wojtak} {et~al.}(2013){Wojtak}, {Gottl{\"o}ber}, \&
  {Klypin}}]{WGK13}
{Wojtak}, R., {Gottl{\"o}ber}, S., \& {Klypin}, A. 2013, \mnras, 434, 1576

\bibitem[{{Wojtak} {et~al.}(2007){Wojtak}, {{\L}okas}, {Mamon},
  {Gottl{\"o}ber}, {Prada}, \& {Moles}}]{Wojtak+07}
{Wojtak}, R., {{\L}okas}, E.~L., {Mamon}, G.~A., {et~al.} 2007, \aap, 466, 437

\bibitem[{{Xie} {et~al.}(2017){Xie}, {De Lucia}, {Hirschmann}, {Fontanot}, \&
  {Zoldan}}]{Xie2017}
{Xie}, L., {De Lucia}, G., {Hirschmann}, M., {Fontanot}, F., \& {Zoldan}, A.
  2017, \mnras, 469, 968

\bibitem[{{Zwicky}(1933)}]{Zwicky33}
{Zwicky}, F. 1933, Helvetica Physica Acta, 6, 110

\end{thebibliography}


\end{document}